\documentclass[11pt,a4paper]{article}
\pdfoutput=1
\usepackage{jcappub}

\usepackage{bm}
\usepackage{bbold}
\usepackage{amsbsy}
\usepackage{subfigure}
\usepackage{slashed}
\usepackage{afterpage}
\usepackage{psfrag}
\usepackage{dsfont}

\renewcommand\vec[1]{{\bf #1}}

\newcommand{\be}{\begin{equation}}
\newcommand{\ee}{\end{equation}}
\newcommand{\bea}{\begin{eqnarray}}
\newcommand{\eea}{\end{eqnarray}}

\newcommand{\vv}[2]{ \left( \begin{array}{cc}   #1 \\ #2 \end{array} \right)}
\newcommand{\vvv}[3]{ \left( \begin{array}{cccc}   #1 \\ #2 \\ #3 \end{array} \right)}
\newcommand{\vvvv}[4]{ \left( \begin{array}{cccc}   #1 \\ #2 \\ #3 \\ #4\end{array} \right)}

\renewcommand\({\left(}
\renewcommand\){\right)}
\renewcommand\[{\left[}
\renewcommand\]{\right]}

\newcommand{\GG}{{\sf G}}
\newcommand{\PP}{{\sf P}}
\newcommand{\TT}{{\sf T}}
\newcommand{\MM}{{\sf M}}
\newcommand{\Ss}{{\sf S}}

\newcommand\chit{\chi_{_b}}
\newcommand\chil{\chi_{_\ell}}

\newcommand{\exclude}[1]{}

\definecolor{gre}{rgb}{0,0.4,0.3}

\begin{document}
\subheader{\hfill MPP-2017-100}

\title{Dielectric haloscopes: sensitivity to the axion dark matter velocity}
\author[a]{Alexander~J.~Millar,}
\author[a,b]{Javier~Redondo,}
\author[a]{Frank~D.~Steffen}

\affiliation[a]{Max-Planck-Institut f\"ur Physik (Werner-Heisenberg-Institut),
F\"ohringer Ring 6,\\ 80805 M\"unchen, Germany}

\affiliation[b]{University of Zaragoza, P.\ Cerbuna 12, 50009 Zaragoza, Spain}

\emailAdd{millar@mpp.mpg.de}
\emailAdd{jredondo@unizar.es}
\emailAdd{steffen@mpp.mpg.de}

\abstract{We study the effect of the axion dark matter velocity in the recently proposed dielectric haloscopes, a promising avenue to search for well-motivated high mass ($40-400~\mu$eV) axions. 
We describe non-zero velocity effects for axion-photon mixing in a magnetic field and for the phenomenon of photon emission from interfaces between different dielectric media. As velocity effects are only important when the haloscope is larger than about $20$\% of the axion de Broglie wavelength, for the planned
MADMAX experiment with 80~dielectric disks the velocity dependence can safely be neglected. 
However, an augmented MADMAX or a second generation experiment would be directionally sensitive to the axion velocity, and thus a sensitive measure of axion astrophysics.}

\maketitle

\section{ Introduction}
The nature of dark matter (DM) remains one of the most enduring mysteries in physics. 
One of the theoretically best motivated DM candidates is the axion.
This low mass pseudoscalar boson is required by the Peccei--Quinn (PQ) mechanism 
which provides a possible resolution to the strong CP problem, 
the absence of CP violation in quantum chromodynamics (QCD)~\cite{Peccei:2006as,
Kim:2008hd, Agashe:2014kda}. 
The value of the axion mass $m_a$ for which axions provide all or a significant part of the relic DM density depends on the cosmological history. 
In the scenario in which the PQ symmetry is broken after inflation~\cite{Sikivie:2009fv,Kawasaki:2013ae},
the observed DM density points to a favoured $m_a$ range of 50--200~$\mu$eV, with $m_a\simeq 100~\mu$eV as a preferred value.
While no experiment yet has been able to probe this parameter space~\cite{Hiramatsu:2012gg,Kawasaki:2014sqa},
a new experimental concept has recently been introduced to search for these `high-mass' axions: 
the dielectric haloscope~\cite{TheMADMAXWorkingGroup:2016hpc}.

Axions interact with photons as described by the Lagrangian 
${\mathcal L}_{a\gamma}=g_{a\gamma}{\bf E}\cdot {\bf B}a$, 
with a dimensionful coupling $g_{a\gamma}\sim\alpha/(2\pi f_a)$, 
the fine-structure constant $\alpha$, 
the PQ scale $f_a$, and the electric and magnetic fields ${\bf E}$ and ${\bf B}$, respectively. 
Accordingly, axions mix with photons in the presence of an external magnetic field, which we will denote as ${\bf B}_{\rm e}$.
Even for a strong ${\bf B}_{\rm e}$ field, the mixing is small and leaves two admixtures of axions and photons in the mass basis: 
a mostly axion-like particle, which acquires a small electric and magnetic field, 
and a mostly photon-like particle, which acquires a small axion component.

While the axion-induced electric and magnetic fields are extremely tiny, 
the conversion of axions to photons can be enhanced by providing suitable geometries. 
This has traditionally been attempted via microwave cavities, 
which would allow the axion field to drive resonant modes of the cavity~\cite{Sikivie:1983ip}. 
By using a cavity with a high quality factor ($Q$-factor) and a large volume $V$ inside a strong ${\bf B}_{\rm e}$ field, 
one expects an enhanced conversion of axions to photons,
as exemplified by the axion DM search experiments ADMX and HAYSTAC \cite{Rybka:2014xca,Kenany:2016tta}. 
However, since the dimensions of the cavity must be on the order of $\lambda/2=0.62~{\rm cm}\, (100~\mu{\rm eV}/m_a)$, 
where $\lambda$ is the Compton wavelength of the axion, cavity haloscopes become increasingly ineffective towards high values of the axion mass. Recently there is increasing interest in building effective cavities operating at higher frequencies~\cite{Rybka:2014cya,Goryachev:2017wpw,McAllister:2017ern,McAllister:2017lkb}. 

\begin{figure}[htbp]
\begin{center}
\includegraphics[width=10cm]{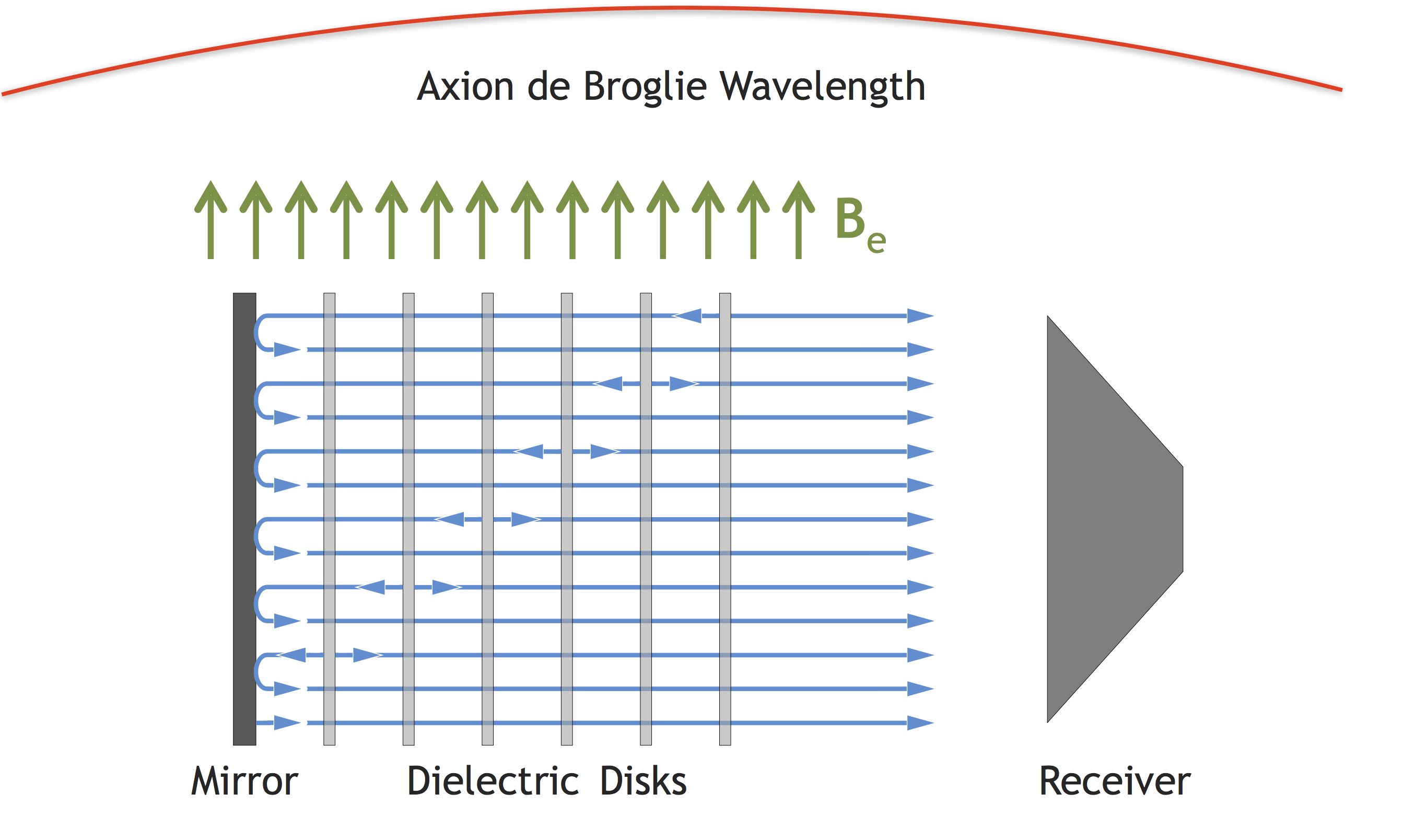}
\caption{Schematic of a dielectric haloscope. In the presence of an axion and a strong external magnetic field ${\bf B}_{\rm e}$, EM waves (blue lines) are emitted from each dielectric disk (gray bars), and then reflect and transmit through the device till they reach the receiver. The mirror (dark gray) ensures that the entire signal power is directed towards the receiver. For simplicity we have not depicted the reflections. A fraction of the axion de Broglie wavelength is also depicted: the effects of the changing phase of the axion throughout the device will be the primary concern of this paper. Figure adapted from \cite{TheMADMAXWorkingGroup:2016hpc} with permission.}
\label{fig:haloscope}
\end{center}
\end{figure}

Rather than using a resonant cavity, a dielectric haloscope uses a series of dielectric disks with adjustable spacings
placed in a strong ${\bf B}_{\rm e}$ dipole field~\cite{TheMADMAXWorkingGroup:2016hpc}. 
In such a setup, axions lead to the generation of electromagnetic (EM) power according to the following principle:
if one places an interface between two different dielectric media inside a magnetic field, 
the axion field causes microwave radiation to be emitted from either side of the interface~\cite{Horns:2012jf}. 
In essence, breaking translation invariance allows one to convert axion-like states to photon-like states. 
One can enhance this effect to an observable level by using multiple interfaces, 
using reflections and constructive interference to boost the signal~\cite{TheMADMAXWorkingGroup:2016hpc}. For an illustration, see figure~\ref{fig:haloscope}. 
Such a dielectric haloscope could be used to search for DM axions in the well-motivated $m_a$ range of $40-400~\mu$eV for the first time. 
The underlying theory of dielectric haloscopes has been explored in detailed studies,~\cite{Millar:2016cjp,Ioannisian:2017srr}, but in these works the velocity of the axion was assumed to be zero.
In the present study we now go beyond the zero-velocity limit and investigate the effects for a velocity of $v \sim 10^{-3}c$,
as expected for galactic cold DM axions on earth. These effects will come almost entirely from the velocity of the axion in the direction perpendicular to the disks, which provides a change in phase of the axion over the haloscope. %
We also discuss strategies to probe the galactic DM velocity distribution in the event of a discovery.

This paper is organised as follows.
We first derive the mixing between axions and photons in matter 
for arbitrary values of the axion velocity and general ${\bf B}_{\rm e}$ alignments in section~\ref{mixing}. 
We then consider the radiation caused by a single dielectric interface inside a magnetic field in a 3D~setting in section~\ref{3D}. 
In section \ref{multilayer} we generalise the transfer matrix formalism developed in the zero-velocity limit in reference~\cite{Millar:2016cjp} 
to allow for a description of the more realistic case with a non-zero axion velocity.
Using this formalism, we study both analytically treatable special cases 
and realistic examples of configurations of dielectric haloscopes in section~\ref{analytic}. 
This includes for the first time a full and realistic 80~disk example of a dielectric haloscope, 
as is being considered for the proposed MADMAX experiment~\cite{TheMADMAXWorkingGroup:2016hpc}. 
From these examples we conclude that the velocity of the axion is only relevant when the haloscope is a large fraction ($\sim20$\%) of the axion's de Broglie wavelength. As such, it can be safely neglected in the MADMAX context. However, as discussed in section~\ref{vdist} if the axion were to be discovered, one could design a larger device capable of measuring the axion's velocity distribution with directional sensitivity.
We reach our conclusions in section~\ref{conclusion}.

\section{Axion-photon mixing in an external magnetic field}
\label{mixing}
Before exploring the effect of the axion's velocity in a complicated setup such as a dielectric haloscope, we must know how the axion mixes with the photon in an external magnetic field~${\bf B}_{\rm e}$. Here we consider the linearised axion-Maxwell equations in the picture of axion-photon mixing. 
\subsection{Equations of motion}

The interaction between photons and axions is described by the Lagrangian density
\begin{equation}\label{lagrangian}
{\cal L} = -\frac{1}{4}F_{\mu\nu}F^{\mu\nu}-J^\mu A_\mu+\frac{1}{2}\partial_\mu a \partial^\mu a
-\frac{1}{2}m_a^2a^2 -\frac{g_{a\gamma}}{4}F_{\mu\nu}\widetilde F^{\mu\nu}a, 
\end{equation} 
with $a$ the axion field with mass
$m_a$. The coupling strength between axions and photons is governed by $g_{a\gamma}$.
We use $F_{\mu\nu}=\partial_\mu A_\nu-\partial_\nu A_\mu$ to denote the EM
field-strength tensor in terms of the vector potential
$A^\mu=(A_0,{\bf A})$ and $J^\mu=(\rho,{\bf J})$ is the electric 4-current 
involving the charge density $\rho$ and the current density ${\bf J}$. The dual tensor is given by
\smash{$\widetilde F^{\mu\nu} = \frac{1}{2}\varepsilon^{\mu\nu\alpha\beta}F_{\alpha\beta}$}, where we use the Levi-Civita tensor with $\varepsilon^{0123}=\varepsilon_{123}=+1$. The electric and
magnetic fields are given by
\begin{equation}
{\bf E} = -{\bm \nabla} A_0-\dot {\bf A}
\qquad\hbox{and}\qquad
{\bf B} =  {\bm \nabla}\times  {\bf A}\,.
\end{equation}
Throughout this paper we will use natural units with $\hbar = c = 1$ and the Lorentz-Heaviside convention
$\alpha=e^2/4\pi$, giving
$1~{\rm V}/{\rm m}=6.5163\times 10^{-7}~{\rm eV}^2$ and
$1~{\rm T}=1~{\rm Tesla}=195.35~{\rm eV}^2$. The energy density of the EM field
is $\frac{1}{2}({\bf E}^2+{\bf B}^2)$. 

While the mass and photon couplings are independent for a generic axion-like particle (ALP), we
are mostly interested in QCD axions. In this case, both parameters are largely decided by
the PQ scale, or axion decay constant, $f_a$ by
$m_af_a\sim m_\pi f_\pi$ and $g_{a\gamma}\sim\alpha/(2\pi f_a)$, where $m_\pi$ and
$f_\pi$ are the pion mass and decay constant, respectively. The most recent study
gives the numerical values \cite{diCortona:2015ldu}
\begin{subequations}
\begin{eqnarray}
m_a&=&5.70(6)(4)\, {\rm \mu eV}\,\left(\frac{10^{12}\rm\,GeV}{f_a}\right)\,,\label{eq:ma}\\
g_{a\gamma}&=&-\frac{\alpha}{2\pi f_a}\,C_{a\gamma}
=-2.04(3)\times10^{-16}~{\rm GeV}^{-1}\,\left(\frac{m_a}{1\,\mu{\rm eV}}\right)\,C_{a\gamma}\,,\label{eq:gag}\\
C_{a\gamma}&=&\frac{{\cal E}}{{\cal N}}-1.92(4)\label{eq:cag}\,,
\end{eqnarray}
\end{subequations}
where the numbers in brackets denote the uncertainty in the last digit,
with the first and second error given for $m_a$ referring to quark-mass uncertainties and higher order corrections, respectively. $C_{a\gamma}$ is a model dependent number of ${\cal O} (1)$, which is given by the EM and colour anomalies (${\cal E}$ and $\cal N$, respectively). In models where ordinary quarks and leptons do not carry PQ charges,
the axion-photon interaction arises
entirely from $a$-$\pi^0$-$\eta$ mixing and ${{\cal E}}/{{\cal N}}=0$,
the KSVZ model \cite{Kim:1979if, Shifman:1979if}
being the traditional example. In more general models,
${{\cal E}}/{{\cal N}}$ is a ratio of small integers, the DFSZ model \cite{Dine:1981rt, Zhitnitsky:1980tq}
with ${{\cal E}}/{{\cal N}}=8/3$ being
an often-cited example, although there exist many other cases \cite{Kim:2014rza}. Recent studies have explored the range of $C_{a\gamma}$, finding that for the QCD axion $0\lesssim C_{a\gamma}\lesssim 55$, though the extreme cases require judicious parameter choices~\cite{DiLuzio:2016sbl,DiLuzio:2017pfr}. ALPs can have any value of $C_{a\gamma}$ not ruled out by experiments or observations.

From \eqref{lagrangian} we can derive the axion-Maxwell equations in terms of electric and magnetic fields, \cite{Sikivie:1983ip, Wilczek:1987mv}
\begin{subequations}
\begin{eqnarray}
{\bm\nabla}\cdot {\bf E} &=& \rho -	g_{a\gamma} {\bf B}\cdot{\bm\nabla} a\,,
\label{eq:Maxwell-a}\\
{\bm\nabla}\times {\bf B}- \dot{\bf E}  &=& {\bf J}
+g_{a\gamma}\({\bf B}\, \dot a -{\bf E}\times{\bm\nabla} a\)\,,\label{eq:Maxwell-b}\\
{\bm\nabla}\cdot{\bf B}&=& 0\,,\\
{\bm\nabla}\times{\bf E}+\dot{\bf B}&=&0\,,\label{eq:Maxwell-d}\\
\ddot a-{\bm\nabla}^2 a +m_a^2 a &=&
g_{a\gamma} {\bf E}\cdot {\bf B}\,.\label{eq:Maxwell-c}
\end{eqnarray}
\end{subequations}
While the homogenous equations do not involve the axion field, they will be
used to set the EM boundary conditions for an
interface. 

\subsection{Linearized axion-Maxwell equations}
\label{linearization}
Our situation of interest is one where a strong static external magnetic field
${\bf B}_{\rm e}$ has been set up by means of an external current ${\bf J}_{\rm e}$
and is used to mix axions (in our case DM axions) with EM waves. As shown explicitly, e.g., in Ref.~\cite{Millar:2016cjp}, 
such a setting can be described by linearised macroscopic axion-Maxwell equations, which have plane-wave solutions. This description in terms of linear particle mixing is possible due to the significant hierarchy of scales between the static ${\bf B}_{\rm e}$ field, the background axion field and the induced EM fields, i.e., ${\bf B}_{\rm e}$ is much larger than all other fields, and the background axion field is much larger than any induced EM fields. 
 For related linearisations, see also~\cite{Wilczek:1987mv,Das:2004qka,Das:2004ee,Ganguly:2008kh,VISINELLI:2013fia}.
Expanding the time-dependent fields in plane waves proportional to $\exp\[-i(\omega t - {\bf k}\cdot{\bf x})\]$ leads then
to the linearised macroscopic axion-electrodynamic equations in Fourier space~\cite{Millar:2016cjp}
\begin{subequations}
\begin{eqnarray}
\epsilon{\bf k}\cdot {\bf E} &=& -g_{a\gamma}{\bf k}\cdot{\bf B}_{\rm e}a\,,
\label{eq:Maxwell-aaa}\\
{\bf k}\times {\bf B}/\mu+\omega \epsilon{\bf E}  &=&
-g_{a\gamma}\omega{\bf B}_{\rm e} a\,,\label{eq:Maxwell-bbb}\\
{\bf k}\cdot{\bf B}&=& 0\,,\\
{\bf k}\times{\bf E}-\omega{\bf B}&=&0\,,\\
\(\omega^2-{\bf k}^2-m_a^2\) a &=&
-g_{a\gamma} {\bf B}_{\rm e}\cdot{\bf E}\,,\label{eq:Maxwell-ccc}
\end{eqnarray}
\label{eq:Maxwell-macroscopic}
\end{subequations}
where ${\bf E}$, ${\bf B}$, $a$, $\epsilon$ and $\mu$ depend on $\omega$ and ${\bf k}$. In the macroscopic equations~\eqref{eq:Maxwell-macroscopic}, the media is assumed to be homogeneous and isotropic,
with a linear response that can be encoded 
for the electric displacement field ${\bf D}=\epsilon{\bf E}$ in terms of the dielectric constant $\epsilon$
and for the magnetic field ${\bf H}={\bf B}/\mu$ in terms of the magnetic permeability $\mu$.
Notice that
in our natural units, $\epsilon$ and $\mu$ in vacuum are equal to unity,
i.e., they are what is usually called the {\em relative\/} dielectric
permittivity and {\em relative\/} magnetic permeability. In lossy media, $\epsilon$ and $\mu$ have imaginary parts.

\begin{figure}[t]
\begin{center}
\includegraphics[width=7cm]{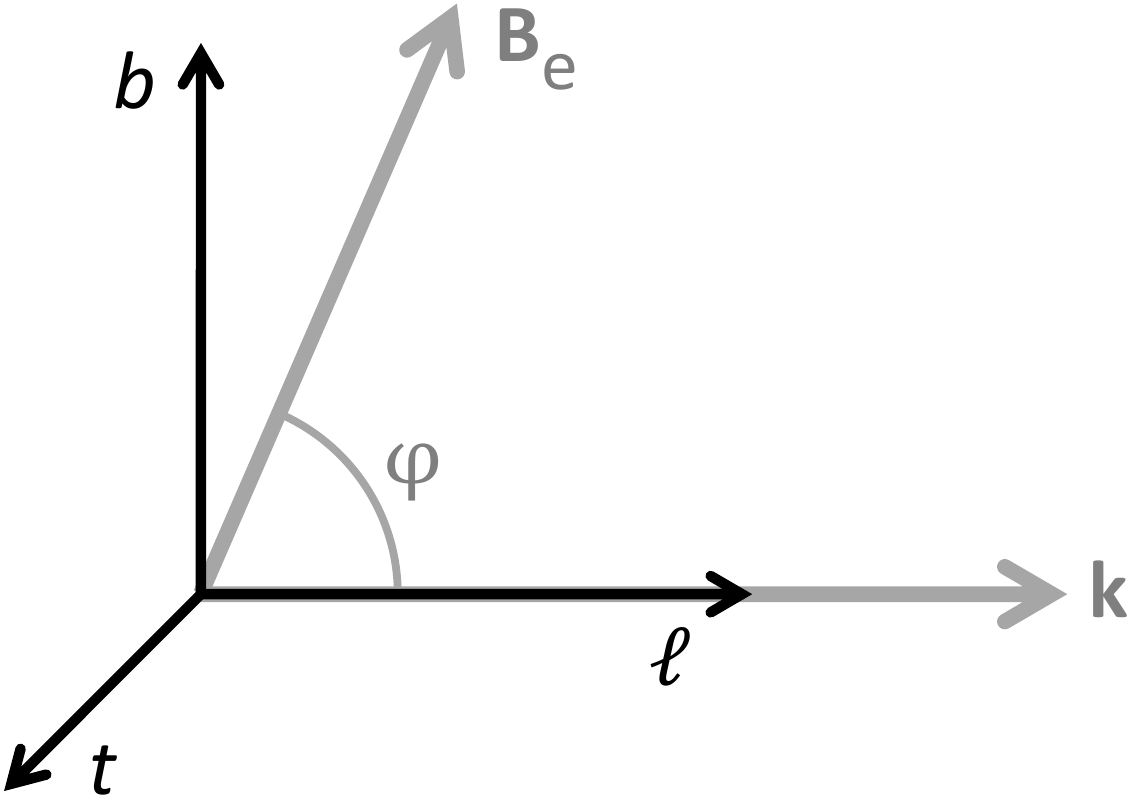}
\caption{Polarization basis for a field with wavevector ${\bf k}$,
defining the $\ell$ direction. Perpendicular to ${\bf k}$ and
${\bf B}_{\rm e}$ is the \emph{t} direction, whereas the \emph{b} direction is
perpendicular to ${\bf k}$ and spans the plane containing ${\bf B}_{\rm e}$.}
\label{axes}
\end{center}
\end{figure}

We gain more insight by expressing ${\bf E}$ and ${\bf B}$ in terms of
their components along ${\bf k}$, which we call the $\ell$ (longitudinal)
component, and the transverse components. To describe the latter, we
use a coordinate system with one vector in the plane of ${\bf k}$ and
${\bf B}_{\rm e}$, which we call the \emph{b} direction, and one vector
transverse to both ${\bf k}$ and ${\bf B}_{\rm e}$, which we
call the \emph {t} direction, (see figure~\ref{axes}). In our practical applications,
typically ${\bf B}_{\rm e}$ will be very nearly perpendicular to
${\bf k}$ so that the ``\emph{b} direction'' then almost coincides with ${\bf B}_{\rm e}$.
We denote the angle between ${\bf B}_{\rm e}$ and ${\bf k}$ as $\varphi$ (as in figure~\ref{axes}) so that the external $B$-field in the
$b$ direction is $B_{\rm e}\sin\varphi$ and the one in the $\ell$ direction
is $B_{\rm e}\cos\varphi$.

Equation~(\ref{eq:Maxwell-aaa}) applies to the $\ell$ component and simply
reads $k\epsilon E_{\rm \ell}=-g_{a\gamma}k B_{\rm e}a\cos{\varphi}$, whereas
the $\ell$ part of equation~(\ref{eq:Maxwell-bbb}) reads
$\omega\epsilon E_{\rm \ell}=-g_{a\gamma}\omega B_{\rm e}a\cos{\varphi}$. These
equations are equivalent except when $\omega=0$ or $k=0$. We only
consider solutions with $\omega\not=0$ because by assumption the only static field
is ${\bf B}_{\rm e}$.
On the other hand, we also consider homogeneous situations ($k=0$), so overall
we may drop equation~(\ref{eq:Maxwell-aaa}), which is trivial when $k=0$, and keep the $\ell$ part of
equation~(\ref{eq:Maxwell-bbb}). 

It is convenient to return to the vector potential as a description of the
time-varying ${\bf E}$ and ${\bf B}$ fields. In Fourier space we have
${\bf B}=i\,{\bf k}\times{\bf A}$ and ${\bf E}=i(\omega{\bf A}-{\bf k}A_0)$.
In the longitudinal components of the equations,
only the combination $\omega A_{\rm \ell}-k A_0$ ever appears. Because in our case
$\omega\not=0$, it is consistent to set $A_0=0$ which amounts to using
the temporal gauge (also known as Weyl or Hamiltonian gauge). The
subtleties of this choice do not affect our case. Alternatively, because $\omega\not=0$,
we could define $A_{\rm \ell}$ simply as a notation for the
physical quantity $E_{\rm \ell}/i\omega$.
In terms of ${\bf A}$, the remaining linear equations of motion are 
\begin{subequations}
\begin{eqnarray}
\(\epsilon\,\omega^2-{\bf k}^2/\mu\){\bf A}+({\bf k}\cdot{\bf A}){\bf k}/\mu
&=&ig_{a\gamma}\omega{\bf B}_{\rm e}\,a\,,\label{eq:Maxwell-b4}\\
\(\omega^2-{\bf k}^2-m_a^2\)\,a &=&
-ig_{a\gamma}\omega{\bf B}_{\rm e}\cdot{\bf A}\,.\label{eq:Maxwell-c4}
\end{eqnarray}
\end{subequations}
In terms of the different polarization components, our final equations
of motion in Fourier space are
\begin{equation}\label{mixingM}
\vvvv{\omega^2-k^2-m_a^2 &\Delta_{\rm \emph b}\omega & \Delta_{\rm \ell}\omega&0}
{\Delta_{\rm \emph b}\omega & \epsilon\omega^2-k^2/\mu & 0&0}
{\Delta_{\rm \ell}\omega & 0 & \epsilon\omega^2&0}{0 & 0 &0&\epsilon\omega^2-k^2/\mu}
\vvvv{a}{iA_{\rm \emph b}}{iA_{\rm \ell}}{iA_{\rm  \emph t}}= 0\,,
\end{equation}
where the ``mixing energies'' are $\Delta_{\rm \emph b}=g_{a\gamma}
B_{\rm e}\sin{\varphi}$
and $\Delta_{\rm \ell}=g_{a\gamma} B_{\rm e}\cos{\varphi}$. We recall
that in general the response functions
$\epsilon$ and $\mu$ depend on $\omega$ and ${\bf k}$
and that in lossy media they have imaginary parts. The factor of $i$ in \eqref{mixingM} reflects a phase difference of $\pi/2$ between the components of $\bf A$ and $a$, which disappears when $\bf A$ is expressed in terms of $\bf E$.

\subsection{Eigenmodes of the interacting system}

The last line of the equations of motion (\ref{mixingM}) remains
unaffected by the axion-photon interaction and leads to the
dispersion relation $k^2=n^2\omega^2$, where we have introduced the usual
refractive index\footnote{We assume that the
imaginary part of $n$ is much smaller than its real part
and that the real part is positive. In this
case we can use relations of the type $x=\sqrt{x^2}$
with impunity because
we avoid the branch cut of the square-root function along the
negative real axis.}
\begin{equation}
n^2={\epsilon\mu}\,.
\end{equation}
Notice that in general $n$ can be a complicated function of $\omega$ and $k$,
although in our cases of practical interest, we ignore spatial dispersion
(no dependence on $k$) and  temporal dispersion (the dependence on $\omega$), as we expect them to be weak. The same
dispersion relation applies, of course, to the \emph b polarization in the absence
of axion-photon mixing. In this case the $\ell$ dispersion relation is
$\epsilon\omega^2=0$ and, because we only consider $\omega\not=0$, would require $\epsilon({\omega,k})=0$ to support a propagating mode. A plasma can provide such a mode,
the longitudinal plasmon. A non-relativistic plasma has $\epsilon\simeq(1-\omega_p^2/\omega^2)$, where $\omega_p$ is the plasma frequency (i.e., the frequency of the longitudinal plasmon). While in principal axions can mix resonantly with longitudinal plasmons~\cite{Das:2004qka}, there are currently no proposed experimental efforts using such a mixing. The dielectric media we will consider later do not support such
dynamical $\ell$ modes. The axion dispersion relation in the absence of a magnetic field is $\omega^2-k^2=m_a^2$.

In the presence of axion-photon interactions, the eigenmodes of our system are
superpositions of the axion and photon field. Note that $A_{\rm t}$ seems to decouple in equation \eqref{mixingM} but when we consider a boundary between different dielectric media we will see that non-trivial alignments between the interface and the magnetic field can mix $A_{\rm t}$ with $A_{\rm b}$.

Nontrivial solutions of equation~(\ref{mixingM}) require the
determinant of the matrix to vanish. Because in general $\epsilon$
and $\mu$ depend on both $\omega$ and ${\bf k}$, the solution
depends on the material properties of the medium. Here we will neglect spatial dispersion, i.e., henceforth we assume that $\epsilon$ and $\mu$ depend
only on $\omega$. Moreover, we also assume isotropic media where
the response functions do not depend on the direction of ${\bf k}$.
We then find two solutions for the dispersion relation.
To lowest non-trivial order in the extremely small mixing energies they are
\begin{subequations}\label{eq:dispersionrelations}
\begin{eqnarray}
\hbox to 6em{Axion like: \hfil}~k^2&=&  \omega^2-m_a^2- \frac{\Delta_b^2\omega^2\mu }{n^2\omega^2-\omega^2+m_a^2}
-\frac{\Delta_\ell^2\omega^2\mu}{n^2\omega^2}+{\cal O}(\Delta_{b,\ell}^4)\,\\
\hbox to 6em{Photon like:\hfil}~k^2&=&n^2\omega^2
+ \frac{\Delta_b^2\omega^2\mu }{n^2\omega^2-\omega^2+m_a^2}+{\cal O}(\Delta_{b,\ell}^4)\,.
\end{eqnarray}
\end{subequations}
A resonance can occur when $n<1$, i.e., when the ``squared effective photon mass''
$m_\gamma^2=\omega^2-k^2=(1-n^2)\omega^2$ is positive. However, as the mixing is typically exceedingly small, even tiny losses in the medium will limit the resonance enough so that higher order terms are not required. For example, for 100~$\mu$eV DM axions in a 10~T $B$-field $\Delta_{b,\ell}\sim 10^{-12}~\mu$eV, so ${\rm Im}( n)\lesssim 10^{-26}$ is required for $\Delta_{b,\ell}^4$ effects to become relevent in \eqref{eq:dispersionrelations}. Further, to get the resonance condition one would need $n\sim 10^{-3}$, so an incredibly special material would be needed to realise this. For higher order terms to be necessary, the numerator and denominator must be of similar size, which requires very large frequencies, low losses and large mixing energies. Such resonances can occur in the often-studied case of axion-photon oscillations \cite{Raffelt:1987im} at
high energies, where one typically assumes axions and photons
to be nearly degenerate, i.e., it is assumed that
\smash{$\Delta_b\omega\gg m_a^2$} and \smash{$\Delta_b\omega\gg |n^2-1|\omega^2$}.
Actually at the highest astrophysically relevant energies in the
10~TeV range, intergalactic magnetic fields provide this condition as the cosmic microwave background then produces
a non-negligible refractive index~\cite{Dobrynina:2014qba}.  

Motivated by our case of an ordinary dielectric medium,
we have assumed a
photon refractive index of $n>1$, i.e., the photon dispersion relation is
space like, ignoring a possible small imaginary part.
In this case
the two branches of the
dispersion relation do not cross for any value of $\omega$; photons
and axions are never on resonance with each other, and so the
denominator in \eqref{eq:dispersionrelations} never becomes zero. Note that the velocities of the particles are given by the group velocity of the waves, $\partial\omega/\partial k$. For the axion this then gives us that ${\bf v}={\bf k}/\omega$

It is now straightforward to obtain the field configurations
corresponding to these propagating modes. To lowest order in the small
mixing energies we find
\begin{subequations}
\begin{eqnarray}
\hbox to 6em{Axion like: \hfil}~\vvv{a}{i A_{\emph b}}{iA_{\ell}} &=&
\vvv{1}{-\chit \mu \sin\varphi}{-\chil \mu \cos\varphi} +{\cal O}(\chi_{ b,\ell}^2),\\[1ex]
\hbox to 6em{Photon like:\hfil}~\vvv{a}{i A_{\emph b}}{iA_{\ell}}&=&
\vvv{\chit \sin\varphi}{1}{0} + {\cal O}(\chi_{ b,\ell}^2)\,,
\end{eqnarray}
\end{subequations}
where we have defined the small dimensionless mixing parameters
\begin{subequations}
\begin{eqnarray}
\chit &=& \frac{g_{a\gamma} B_{\rm e}\,\omega}{n^2\omega^2-\omega^2+m_a^2}\,, \\
\chil &=& \frac{g_{a\gamma} B_{\rm e}\,\omega}{n^2\omega^2}\,.
\end{eqnarray}
\end{subequations}
These vectors are normalized at linear order in $\chi_{b,\ell}$ (linear order
in $g_{a\gamma}$). On the other hand, the dispersion relations in
equation~(\ref{eq:dispersionrelations}) get modified only at order
$g_{a\gamma}^2$. Therefore, the dominant effect of the axion-photon interaction
is that the normal propagating photon and axion modes obtain a small admixture of the other flavor. Henceforth we will always work to linear order
in $g_{a\gamma}$ and thus ignore the modification of the dispersion relations.

Notice that these two solutions correspond to
different combinations of $\omega$ and $k$, i.e., those which obey
the respective dispersion relations. For a fixed choice
of $\omega$, these two solutions have different $k$ values. 

Recall that in our
dielectric media, there is no longitudinal photon-like solution.
Regardless, through the mixing the axion-like mode carries a small electric field component
along ${\bf k}$.\footnote{Likewise, we could consider frequencies below the axion mass, $\omega< m_a$, where the axion has
no propagating mode, yet the photon-like mode exists and
still carries a small axion component.} This small longitudinal electric field is often neglected in the literature---while for most experimental concepts the $B$-field is transverse by design, there are exceptions such as the AMELIE helioscope proposal~\cite{Galan:2015msa}. As in this case the $B$-field is not aligned with the solar axions, it is possible that longitudinal axion induced $E$-fields could influence the experiment. We will show in section \ref{3D} that the axion induced longitudinal field can influence the generation of photon-like waves from an interface.

\subsection{Dark matter axions}

Our case of interest is very special in that the galactic DM axions
are highly non-relativistic ($v\lesssim 10^{-3}$). In view of the small axion mass and velocity, and concomitant large de Broglie wavelength,
\begin{equation}
	\lambda_{\rm dB}
	=\frac{2\pi}{m_a v}
	=12.4~{\rm m}~\left(\frac{100~\mu{\rm eV}}{m_a}\right)\left(\frac{10^{-3}}{v}\right),
\label{eq:lambdadeBroglie}
\end{equation}
the axion field only exhibits a slow change of phase over the laboratory scale. As explained earlier, we ignore a frequency shift by
axion-photon mixing which is quadratic in the small coupling constant $g_{a\gamma}$.
The difference between the transverse and longitudinal mixing is negligible,
\begin{equation}
\chit\simeq\chil\equiv\chi=\frac{g_{a\gamma}B_{\rm e}}{n^2 m_a}
=3.98(5)\times 10^{-16}~\frac{C_{a\gamma}}{n^2}~\frac{B_{\rm e}}{10\,\rm T}\,.
\end{equation}
 The axion-like wave is given by 
\be
a(t,{\bf x}) = a({\bf p}) e^{i({\bf p}\cdot {\bf x}-\omega t)} \quad ; \quad
i{\bf A}_a(t,{\bf x}) =  -  \chi \mu {\bf \hat B}_e\, a(t,{\bf x}) ,\label{dmwave}
\ee
where ${\bf A}_a$ is the axion induced ${\bf A}$ field.
We see that the axion field develops a small electric field oscillating with the axion's frequency $\omega$,
\begin{equation}\label{eq:Ea}
{\bf E}_a(t,{\bf x})= -\dot {\bf A}_a(t,{\bf x})=-\chi\mu\omega\,a(t,{\bf x}){\bf \hat B}_{\rm e}=-\frac{g_{a\gamma}{\bf B}_{\rm e}}{\epsilon}\,a(t,{\bf x})\,.
\end{equation}
In addition, there is also a small induced $H$-field given by
\be
\label{eq:Ha}
{\bf H}_a(t,{\bf x}) = \frac{1}{\mu}\nabla\times {\bf A}_a(t,{\bf x}) = - v \sin\varphi\, \chi \,  \omega  a(t,{\bf x}) {\bf \hat t}_a=-\frac{vg_{a\gamma}B_{\rm e}\sin\varphi}{\mu\epsilon}a(t,{\bf x}) {\bf \hat t}_a.
\ee
Note that this is suppressed by the axion velocity; the axion induced magnetic field will only have a subdominant effect in dielectric haloscopes. However, there are proposals to use this field to detect DM axions by using an LC circuit~\cite{Sikivie:2013laa}. 
The photon-like waves are very much like ordinary transverse EM waves (transverse up to $O(\chi^2)$ corrections) accompanied by a relativistic axion field proportional to the ${\bf \hat b}$ projection of the magnetic field, i.e., ${\bf \hat B}\cdot {\bf \hat b}=\sin\varphi$ (with $\varphi$ defined as in figure~\ref{axes})
\be
{\bf A}(t,{\bf x})  = A_0{\bf \hat b} e^{i({\bf k}\cdot {\bf x}-\omega t)} \quad ; \quad
a_{\gamma}(t,{\bf x}) = i  \chi  {\bf \hat B}_e \cdot {\bf A}(t,{\bf x}) =
\frac{\chi}{\omega}  {\bf \hat B}_e \cdot {\bf E}(t,{\bf x}). \label{photonlike}
\ee
with $n^2\omega^2=|{\bf k}|^2$. As we will be converting axions to photons, all photon-like waves will be produced at linear order in $\chi$, so the axionic component is ${\cal O}(\chi^2)$ and thus negligible.

Apart from an overall phase, the local axion field is 
\be
a(t,{\bf x})= \int  \frac{d^3p}{(2\pi)^3}a({\bf p})e^{-i[\omega t-{\bf p}\cdot {\bf x}  +\alpha({\bf p})]},
\ee
where $\alpha({\bf p})$ is a relative phase for each momenta. The axion field strength $a({\bf p})$ in general can give a non-trivial dependence on ${\bf p}$. The local axionic DM density in some large averaging volume $V$ is
\begin{equation}
\rho_a= \frac{1}{V}\int \frac {d^3p}{(2\pi)^3} \frac{\omega^2 |a({\bf p})|^2}{2}\simeq\frac{1}{V}\int \frac {d^3p}{(2\pi)^3} \frac{m_a^2 |a({\bf p})|^2}{2}=f_{\rm DM}\,\frac{300~{\rm MeV}}{{\rm cm}^3}\,, \label{eq:density}
\end{equation}
where $f_{\rm DM}$ is a fudge factor expressing the uncertainty in the local
dark-matter density\footnote{The local DM density has been estimated by various
authors using different data and assumptions \cite{Catena:2009mf, Strigari:2009zb, Weber:2009pt, Bovy:2012tw, Nesti:2013uwa, Bozorgnia:2013pua, Read:2014qva}. The value
$300~{\rm MeV}~{\rm cm}^{-3}$ is often used as a benchmark number, although
in the axion literature, $400~{\rm MeV}~{\rm cm}^{-3}$ is frequently used. The particle
data group \cite{Agashe:2014kda} gives a even larger value based on Ref.~\cite{Catena:2009mf}.} as well as the uncertainty
of the dark-matter fraction
consisting of axions relative to possible other forms of DM or
relative to the fraction gravitationally bound in axion mini clusters. Note that the factor of $1/2$ in \eqref{eq:density} comes from the cycle average of the axion field.
 We find
\begin{equation}
E_{a}({\bf p})=-\frac{g_{a\gamma} B_{\rm e} a({\bf p})}{\epsilon}=1.3\times10^{-12}~{\rm V}/{\rm m}~\(\frac{B_{\rm e}}{10~{\rm T}}\)~
\frac{C_{a\gamma}f_{\rm DM}^{1/2}}{\epsilon}\frac{a({\bf p})}{a_0}\,
\end{equation}
where $a_0$ would be the field strength if axions were monochromatic with momentum ${\bf p}_0$, i.e., $a({\bf p})=a_0(2\pi)^3\delta^3({\bf p}-{\bf p}_0)$. 
For the QCD axion, this result is independent of the axion mass.

Now that we have found the mixed axion and photon fields, including a non-zero velocity, we can consider a 3D axion field interacting with an interface. From this we will build up a transfer matrix formalism that includes velocity effects.

\section{Axion DM induced boundary EM radiation: 3D calculation with non-zero axion velocities}
\label{3D}
\begin{figure}[tp]
\begin{center}
\includegraphics[width=15.5cm]{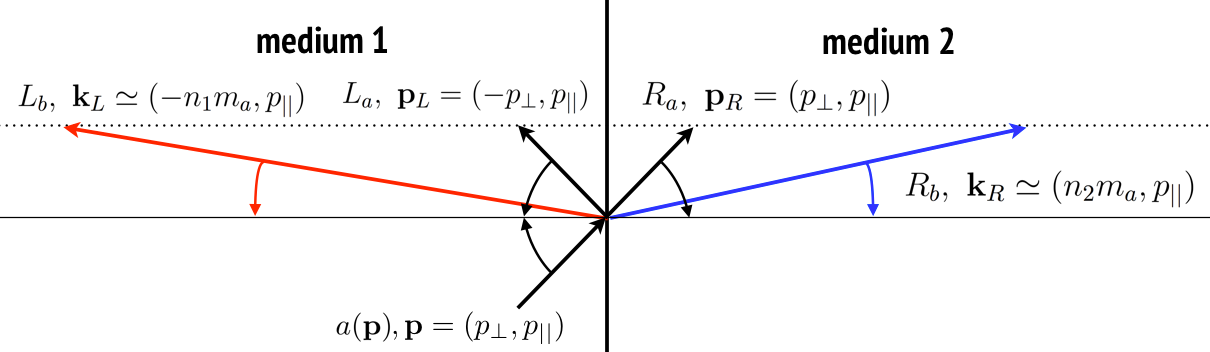}
\caption{\small A non-relativistic axion DM wave of amplitude $a({\bf p})$ and momentum ${\bf p}$ crosses from medium~1 to medium~2 becoming the transmitted right moving wave of amplitude $R_a\simeq a({\bf p})$, with a negligible reflected left moving wave $L_a\simeq 0$. In general the change in media causes a discontinuity in the axions $E$ and $H$ fields that will be compensated by emission of EM waves $L_b,R_b$ (as well as the not pictured $L_t,R_t$). The momentum parallel to the boundary is conserved and the perpendicular momentum is fixed by the dispersion relations, leading the EM waves to be emitted almost perpendicular to the boundary. Different media are those with different axion-photon mixing energies, so any change in the $B$-field intensity, direction or dielectric constant induces radiation of EM waves from both sides of the interface (red and blue arrows). }
\label{medium}
\end{center}
\end{figure}

Now that we have discussed the general axion-photon mixing in matter, we can turn our attention to the effect of a change in media. Here we will consider a fully general 3D case, albeit with an interface of infinite extend, as depicted in figure~\ref{medium}. 
We allow both the axion velocity and the $B$-field to have a non-trivial orientation with respect to the boundary. However, while this situation is technically more complex, as in \cite{Horns:2012jf} the solution still reduces to compensating discrepancies in the axion induced $E$ and $H$ fields with (mostly) regular EM waves. 

The EM and axion fields in medium 1 can be represented as a combination of the incoming axion-like wave, and reflected axion-like and photon-like waves 
\bea
\nonumber
\vvvv{a}{i A_{\emph b}}{iA_{\ell}}{iA_t}_1=&&a({\bf p})\vvvv{1}{-\chi_1\mu_1\sin\varphi}{-\chi_1\mu_1\cos\varphi}{0}e^{i({\bf p}\cdot {\bf x})}+
L_a\vvvv{1}{-\chi_1\mu_1\sin\varphi_{L_a}}{-\chi_1\mu_1\cos\varphi_{L_a}}{0}e^{i({\bf p}_L\cdot {\bf x})}\\
&& + 
L_b\vvvv{\chi_1\sin\varphi_{L_b}}{1}{0}{0}e^{i({\bf k}_{L}\cdot {\bf x})}+L_t\vvvv{0}{0}{0}{1}e^{i({\bf k}_L\cdot {\bf x})}
\eea
while in medium 2 the transmitted axion-like and created photon-like waves are 
\bea
\vvvv{a}{i A_{\emph b}}{iA_{\ell}}{iA_t}_2=&&R_a\vvvv{1}{-\chi_2\mu_2\sin\varphi_{R_a}}{-\chi_2\mu_2\cos\varphi_{R_a}}{0}e^{i({\bf p}_R\cdot {\bf x})}+
R_b\vvvv{\chi_2\sin\varphi_{R_b}}{1}{0}{0}e^{i({\bf k}_R\cdot {\bf x})}\nonumber \\
&&+R_t\vvvv{0}{0}{0}{1}e^{i({\bf k}_R\cdot {\bf x})}.
\eea

We have chosen to write $L_b,~R_b ~(L_t,~R_t)$ to emphasize that we are referring to photon-like waves polarised in the $b~(t)$ direction. Every wave has in principle a different orientation with respect to the external magnetic fields and thus a different $\varphi$ angle (as defined in section~\ref{linearization}). While in this language it seems like $A_t$ plays no role, this is only because we are not in the natural basis of the system, which would be $(x,y,z)$ coordinates.

The boundary conditions on the EM waves are given by
\begin{subequations}
\bea
{\bf E}_{||,1}&=&{\bf E}_{||,2},\\
{\bf H}_{||,1}&=&{\bf H}_{||,2},
\eea\label{eq:boundary}
\end{subequations}
so that without any additional symmetries there are four constraints, requiring four propagating waves to satisfy. The axion field is conserved across the boundary, to good approximation. This will mean that in general both polarisations (which we take to be $A_b, A_t$) will be required in each medium. Choosing $x=x_0$ as our boundary, we first impose
\be
\label{BC1}
\vvv{a}{iA_{b,||}}{iA_{t,||}}_1(t,x_0,y,z) = \vvv{a}{iA_{b,||}}{iA_{t,||}}_2(t,x_0,y,z) , 
\ee
where $A_{b,||},A_{t,||}$ are the projections of the $A_b,A_t$ components in the parallel plane at $x=x_0$. These equations can be only satisfied for all points $(y,z)$ in the boundary if all the phases vary with the same wavenumber along the boundary, i.e., 
\be
{\bf p}_{||}={\bf p}_{||,L}={\bf p}_{||,R}={\bf k}_{||,L}={\bf k}_{||,R} . 
\ee
We can then divide out a common $e^{i {\bf p}_{||}\cdot {\bf x}}$ factor from the boundary condition \eqref{BC1}.
This implies that the angle of the reflected and transmitted photon-like waves with respect to the normal to the surface is suppressed by the small axion DM velocity, 
\be
\sin \kappa = \frac{|{\bf k}_{||}|}{|{\bf k}|} =\frac{|{\bf p}_{||}|}{n \omega} = \frac{|{\bf v}_{||}|}{n} , 
\ee
i.e., photon-like waves are emitted \emph{perpendicularly} to the surface up to ${\cal O}(v/n) \sim 10^{-3}$ corrections. A kin of Snell's law relates the angles of the transmitted and reflected photon-like waves $n_1 \sin \kappa_1=n_2 \sin \kappa_2$. 
For the axion-like waves, the transmission and reflection angles are the same because their dispersion relation is the same in both sides of the boundary, up to $\chi^2$ corrections.

Finally, note that the conservation of momentum and of energy for every wave defines the magnitude of the perpendicular momenta through the dispersion relations. This is particularly important for the photon-like waves, for which $|{\bf k}|^2 = n^2 \omega^2$. This imparts a large perpendicular momentum 
\be
|{\bf k}_{\perp}| = \sqrt{n^2\omega^2-|{\bf k}_{||}|^2} = \sqrt{n^2\omega^2-|{\bf p}_{||}|^2} =  n\omega+\frac{|{\bf p}_{||}|^2}{2 n \omega}+...
=n\omega[1+{\cal O}(v^2)] . 
\ee

In order to continue we make the assumption that the $B$-field \emph{direction} is constant across the boundary. This is not necessary but serves for all the cases we want to cover and simplifies our derivations. We still allow for changes in the magnetic field strength across the boundary. We use ${\bf \hat B}_{\text e}=(\cos\beta,\sin\beta,0)$ with $\sin\beta>0$ in the ($x,y,z$) basis. That is, ${\bf \hat B}_{\text e}$ is in the $x,y$ plane with $z$ being the transverse direction. The $B$-field aligns with the boundary when $\beta=\pi/2$. Note that this assumption of a constant, arbitrary $B$-field direction requires that the $\mu$ of each media is the same. This is also desired for practical reasons, as moving magnetic materials inside strong $B$-fields is practically very difficult. Because of this we set $\mu_r=1$ from here on. 

We also make the stronger assumption that ${\bf B}_{\rm e}$ is homogenous. Inhomogeneities in the $B$-field also would act as a source of momentum for the generated photons~\cite{Redondo:2010js}. We will leave such considerations for future work looking in more depth at this kind of three dimensional effects. In general we will have six boundary conditions, which will require six fields to satisfy them. It is because of this that we must include the transversely polarised photons $A_t$. We can use a change of basis to write the {\bf A} fields associated to the different waves in $(x,y,z)$ coordinates: 
\bea
\label{Afields}
i{\bf A}_{a} &=& -\chi_1 a({\bf p}) (\cos \beta,\sin\beta,0) \\ \nonumber
i{\bf A}_{L_a} &=& -\chi_1 L_a (\cos \beta,\sin\beta,0) \\ \nonumber
i{\bf A}_{R_a} &=& -\chi_2 R_a (\cos \beta,\sin\beta,0) \\ \nonumber
i{\bf A}_{L_b} &=&  \frac{ L_b}{\sin\beta+v_y\cos\beta/n_1}\[(0,\sin\beta,0) +\frac{1}{n_1}(v_y\sin\beta,v_y\cos\beta,v_z\cos\beta)+...\] \\ \nonumber
i{\bf A}_{R_b} &=&\frac{R_b }{\sin\beta-v_y\cos\beta/n_2}\[(0,\sin\beta,0) -\frac{1}{n_2}(v_y\sin\beta,v_y\cos\beta,v_z\cos\beta)+...\]  \\ \nonumber
i{\bf A}_{L_t} &=&  \frac{ L_t}{\sin\beta+v_y\cos\beta/n_1}\[(0,0,-\sin\beta) -\frac{1}{n_1}(v_z\sin\beta,-v_z\cos\beta,v_y\cos\beta)+...\] \\ \nonumber
i{\bf A}_{R_t} &=&\frac{R_t }{\sin\beta-v_y\cos\beta/n_2}\[(0,0,\sin\beta) -\frac{1}{n_2}(v_z\sin\beta,-v_z\cos\beta,v_y\cos\beta)+...\] . \nonumber
\eea
Note that $A_{\ell}$ is a component of ${\bf A}_a$; mismatches in the longitudinal field must also be compensated. In writing the last four equations we have used that 
\begin{subequations}
\bea
{\bf \hat b}_\gamma &=&{\bf \hat t}_\gamma {\mathbf \times \hat \ell}_\gamma,\\
{\bf \hat t}_\gamma &=&-\frac{1}{\sin\varphi_\gamma} {\bf\hat B}_{\text e} \times {\bf \hat \ell}_\gamma, 
\eea
\end{subequations}
with ${\bf \hat \ell}_\gamma=((-1)^{j},v_y/n_j,v_z/n_j)$ for each media $j$. Both photon polarisations have the same~${\bf k}$. Note $\bf\hat b$ is approximately the same for right and left moving waves, whereas from our definition the ${\bf \hat t}$ are almost equal and opposite. To simplify matters, we can express the $\varphi$ angles (that enter into the axion-photon mixing) as
\begin{subequations}
\bea
\sin\varphi_{R_a} &=& \sin\varphi  = ...\\
\cos\varphi_{R_a} &=& \cos\varphi  = ...\\ 
\sin\varphi_{L_b} &=&\sin\varphi_{L_t}= \sin\beta+ \cos\beta \frac{v_y}{n_1} + O(v^2)\equiv \sin\varphi_L \\ \nonumber
\sin\varphi_{R_b} &=& \sin\varphi_{R_t}= \sin\beta- \cos\beta \frac{v_y}{n_2} + O(v^2)\equiv \sin\varphi_R \\
\eea
\end{subequations}
We just need to solve for the values of $L_a,R_a,L_b,R_b,L_t,R_t$ as a function of $a({\bf p})$. 
Explicitly, the boundary conditions \eqref{eq:boundary} are 
\bea
 a(x_0)_1&=&a(x_0)_2,	\\ \nonumber
\partial_x a(x_0)_1 &=& \partial_xa(x_0)_2,	\\ \nonumber
{\bf A}_{y}(x_0)_1 &=&{\bf A}_{y}(x_0)_2,		\\ \nonumber
(\partial_x {\bf A}_{y}-\partial_y {\bf A}_{x})(x_0)_1 &=& (\partial_x {\bf A}_{y}-\partial_y {\bf A}_{x})(x_0)_2,		\\ \nonumber
{\bf A}_{z}(x_0)_1 &=&{\bf A}_{z}(x_0)_2,		\\ \nonumber
(\partial_x {\bf A}_{z}-\partial_z {\bf A}_{x})(x_0)_1 &=&(\partial_x {\bf A}_{z}-\partial_z {\bf A}_{x})(x_0)_2,
\eea
which translate into the following equations at first order in $v$
\be
\small
\kern -2em\left( \begin{array}{ccccccc}   
 1 &  1 & -1 & \chi_1\sin\varphi_{L_b} & -\chi_2\sin\varphi_{R_b}&0&0\\  
v_x &  -v_x & -v_x & -n_1\chi_1\sin\varphi_{L_b} & -n_2\chi_2\sin\varphi_{R_b} &0&0\\  
-\chi_1\sin\beta & -\chi_1 \sin\beta& \chi_2\sin\beta & 1
& -1 & \frac{v_z}{n_1}\cot\beta & -\frac{v_z}{n_2} \cot\beta \\ 
\chi_1g_{-} & \chi_1g_{+} & -\chi_2g_{-} & -n_1
& -n_2 & -v_z\cot\beta & -v_z\cot\beta\\
0 & 0 & 0 & \frac{v_z}{n_1}\cot\beta & \frac{v_z}{n_2} \cot\beta &-1&-1\\ 
v_z\chi_1\cos\beta & v_z\chi_1\cos\beta & -v_z\chi_2\cos\beta & -v_z\cot\beta & v_z\cot\beta  &n_1 &-n_2\\ 
\end{array} \right)
\left( \begin{array}{ccccc}   
a({\bf p}) 
\\  L_a 
\\ R_a 
\\ L_b 
\\ R_b 
\\
L_t 
\\ R_t 
\end{array} \right) = 0,
\ee
where we have taken $x_0=0$ and used ${\bf v}={\bf p}/\omega$. We have defined $g_{\pm}=v_y\cos\beta\pm v_x\sin\beta$.

It is interesting to see that even neglecting the axion components there is an interaction term between the polarisations $b,t$ due to the non-trivial geometry. When these cross terms are non-zero this indicates that both $A_b$ and $A_t$ are required for a consistent solution to the axion-Maxwell equations. One can think of this as coming from the mismatch between the $b,t$ directions and the $y,z$ directions. The full solution is quite involved, but at first order in the tiny quantities $\chi_1, \, \chi_2$ and in the DM velocity $v$ it simplifies to:
\bea
\label{3Dsolution}
L_a &=& 0, \\ \nonumber
R_a &=& a({\bf p}),\\ \nonumber
L_b &=&  a({\bf p}) \(\chi_1-\chi_2\)\sin\beta\frac{n_2}{n_1+n_2}-a({\bf p})\(\chi_1-\chi_2\) \frac{v_x\sin\beta -v_y\cos\beta }{n_1+n_2}, 
\\ \nonumber
R_b &=& - a({\bf p}) \(\chi_1-\chi_2\)\sin\beta \frac{n_1}{n_1+n_2}-a({\bf p})\(\chi_1-\chi_2\)\frac{v_x\sin\beta -v_y\cos\beta}{n_1+n_2},
\\ \nonumber
L_t &=&  L_b\frac{v_z}{n_1}\cot\beta -a({\bf p})\(\chi_1-\chi_2\)\frac{v_z\cos\beta}{n_1+n_2} ,
\\ \nonumber
R_t &=& R_b\frac{v_z}{n_2}\cot\beta+ a({\bf p})\(\chi_1-\chi_2\)\frac{v_z\cos\beta}{n_1+n_2} , 
\eea   
which agrees with the results of \cite{Millar:2016cjp} when $v\to 0$ and $\sin\beta=1$. 
While these solutions are not as elegant as in the $v=0$ case, the interpretations are just as straightforward. The first term in $L_b$ and  $R_b$ is due to the mismatch in the axion-like $E$-field, whereas the second terms in $L_b$,  $R_b$, as well as both terms in $L_t$ and $R_t$ come from a mismatch in the axion-like $H$-field. This can be seen more explicitly by rewriting the last four equations in terms of $E$ and $H$:
\bea
\label{3DsolutionEfields2}
E_L^b &=& \hphantom{+} ({\bf E}_2^a-{\bf E}_1^a)_{y}\frac{n_2}{n_1+n_2}- ({\bf H}_2^a-{\bf H}_1^a)_z\frac{1}{(n_1+n_2)}, 
\\ \nonumber
E_R^b &=& - ({\bf E}_2^a-{\bf E}_1^a)_{y} \frac{n_1}{n_1+n_2}-({\bf H}_2^a-{\bf H}_1^a)_z\frac{1}{(n_1+n_2)},
\\ \nonumber
E_L^t &=& \hphantom{+} ({\bf H}_2^a-{\bf H}_1^a)_y\frac{n_2-n_1}{n_1(n_1+n_2)} ,
\\ \nonumber
E_R^t &=& \hphantom{+} ({\bf H}_2^a-{\bf H}_1^a)_y\frac{n_2-n_1}{n_2(n_1+n_2)} , 
\eea 
where $E_{L,R}^{b,t}$ is the $E$-field traveling in the left and right direction polarised in the $b,t$ direction, respectively. 
As ${\bf B}_{\rm e}$ is defined to lie in the $x,y$ plane there is no mismatch in ${\bf E}_{a,{\bf z}}$ so the axion field does not directly require $A_t\neq0$ as $B_{{\rm e},z}=0$. However, the cross terms between the two photon polarisations $b$ and $t$ do require $A_t$ to be present as well. The symmetry between the two is broken as the axions only interact with $A_b$ (and so the corrections to $A_b$ come at $v^2$ order). 
Maximum emission of the $b$ polarisation waves happens when the $B$-field is exactly along the boundary direction $\bf \hat y$ ($\sin\beta =1$), as noted in \cite{Horns:2012jf}. The amplitude of the waves polarised along $A_t$, i.e., orthogonal to the external $B$-field, are suppressed by $v_z\sim O(10^{-3})$.  While the longitudinal modes $A_{\ell}$ are not dynamical, the mismatch in the axion field induced $A_{\ell,||}^a$ does contribute to the production of photon-like waves. We now have the solution to the 3D boundary case.

\section{Multilayer analysis: dielectric haloscopes}
\label{multilayer}
Given the new dielectric haloscope concept~\cite{TheMADMAXWorkingGroup:2016hpc}, we are interested not only in the production of photons from a single interface, but rather in the cumulative effects of many dielectric layers (as depicted in figure~\ref{fig:haloscope}).
In the presence of several media with numerous interfaces we can arrange them to radiate coherently, enhancing the signal---a dielectric haloscope. With this one could detect axion DM~\cite{TheMADMAXWorkingGroup:2016hpc}. In this section we extend the transfer matrix formalism introduced in \cite{Millar:2016cjp} to include a non-zero axion velocity.
\subsection{Transfer matrix formalism}
\begin{figure}[t]
\begin{center}
\includegraphics[width=12cm]{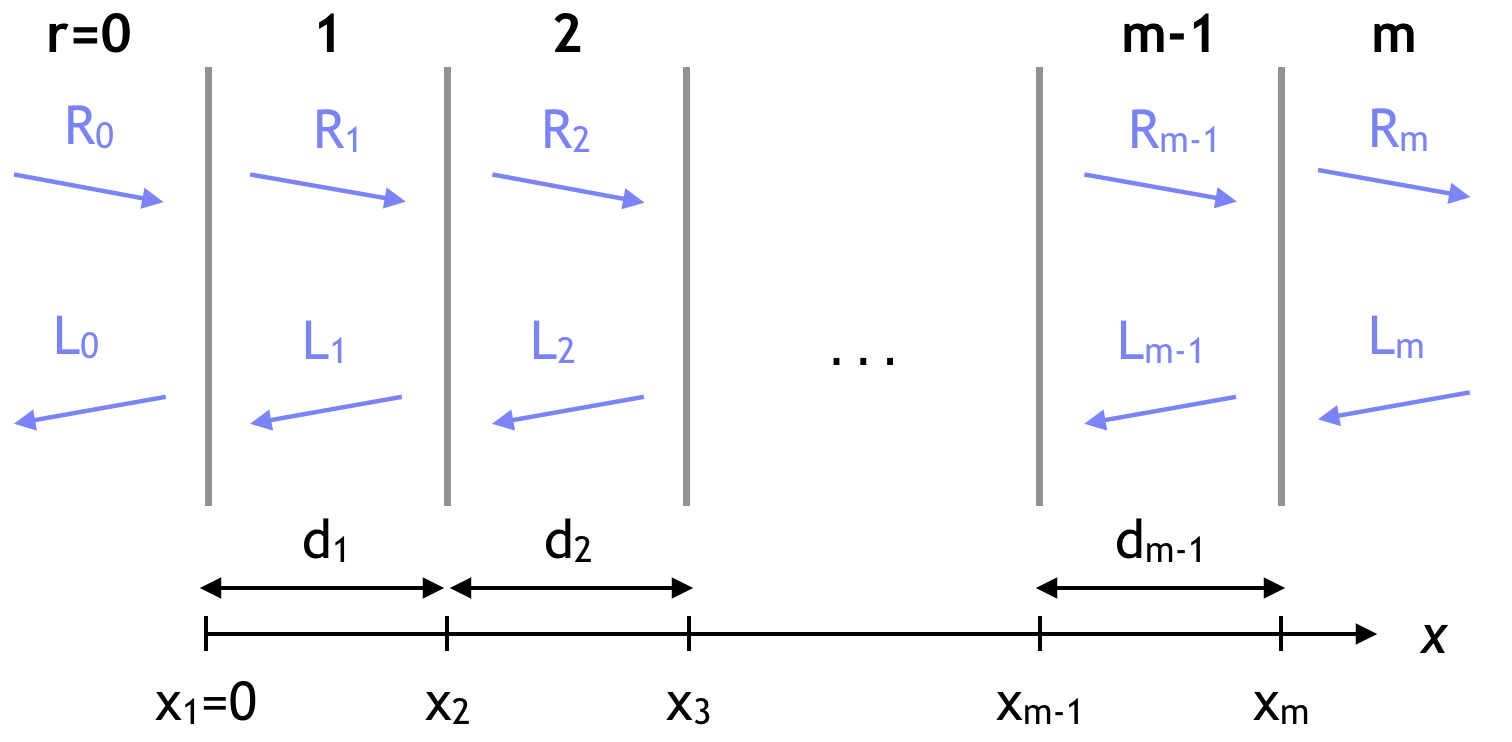}
\caption{Several dielectric regions. The amplitudes $L_r$ and $R_r$ denote
the electric-field amplitudes of left and right moving EM waves in each homogeneous region $r$. The regions end at $x_{r+1}$, covering a distance $d_r$. The angle of photon emission has been exaggerated for visual effect.}
\label{scheme}
\end{center}
\end{figure}

We consider the idealised case of a plane parallel sequence of dielectric regions labelled $0,1... m$, where 0 and m are semi-infinite regions for the output/input, see figure~\ref{scheme}. The thickness of each layer is $d_1,d_2,...,d_{m-1}$ and we allow the indices of refraction to be different ($n_0,...,n_m$) so that the optical thicknesses are $\delta_1=\omega n_1d_1,...$. We consider $\mu=1$ throughout the system, i.e., we assume non-magnetic media.
We consider the external $B$-field aligned with the boundaries, as this corresponds to maximal photon emission, as noted in \cite{Horns:2012jf} and can be seen from taking $\sin\beta=1$ in \eqref{3DsolutionEfields2}. 

Consider an axion DM wave travelling through such a set of dielectric regions. We have learned in the full 3D study of one interface that the axion wave gets transmitted completely through the surfaces, up to higher order effects, and that the momentum of each wave in the direction parallel to the surfaces is conserved in each crossing. As we need to solve for the propagation of the waves generated at each interface via axions, we will need to generalise the transfer matrix formalism used in~\cite{Millar:2016cjp}. The solution of the axion-Maxwell equations is a combination of two plane waves inside each dielectric region together with the axion wave. We have for each region $r$ at first order in the DM velocity ${\bf v}={\bf p}/\omega$ and in $\chi_r$
\be
\vvvv{a}{iA_{x}}{iA_{y}}{i A_z} =
a({\bf p})\vvvv{1}{0}{-\chi_r }{0} 	e^{i v_x \omega x}+
R_r   \vvvv{\chi_r}{-v_y}{1}{0} 						e^{i n_r \omega (x-x_r)}+
L_r    \vvvv{\chi_r}{v_y}{1}{0} 						e^{-i n_r \omega (x-x_r)},
\ee
where we have divided out a common factor of $e^{i(p_y y+p_z z -\omega t)}=e^{i\omega(v_y y+v_z z -t)}$, which does not affect the derived quantities. However, these terms will offset the angle of emitted EM waves.
The continuity of the axion and $\bf E_{||}$-field and its derivatives across the boundary $r,r+1$ imposes in principle four restrictions onto the amplitude of the right and left moving waves. At first order in $\chi\equiv {\rm Max}(\chi_r)$ only the continuity of the $E_{y}$-component and its derivative is relevant as $E_z$ vanishes, giving
\bea
- a({\bf p})e^{iv_x \omega x_{r+1}} \chi_r +R_r e^{i \delta_r} + L_r e^{-i \delta_r} &=& - a({\bf p})e^{iv_x \omega x_{r+1}} \chi_{r+1}+R_{r+1} + L_{r+1},  \\
- a({\bf p})e^{iv_x \omega x_{r+1}} v_x \chi_{r}+n_rR_re^{i \delta_r} - n_r L_re^{-i \delta_r} &=& - a({\bf p})e^{iv_x \omega x_{r+1}} v_x \chi_{r+1}+n_{r+1} R_{r+1} -n_{r+1} L_{r+1} . \nonumber
\eea
 Recalling that $\delta_r=\omega n_r(x_{r+1}-x_r)$, which can be complex if the medium is conducting or absorbing, the system solves to
\bea
\vv{R_{r+1}}{L_{r+1}}&=&
\frac{1}{2n_{r+1}}\(\begin{array}{cc}
n_{r+1}+n_r & n_{r+1}-n_r \\
n_{r+1}-n_r & n_{r+1}+n_r  \end{array}\)
\vv{e^{i \delta_r}R_r}{e^{-i \delta_r}L_r}\nonumber \\
&+& a({\bf p})e^{iv_x \omega x_{r+1}}\frac{\chi_{r+1}-\chi_r}{2} \vv{1+v_x/n_{r+1}}{1-v_x/n_{r+1}} ,
\eea
which we can write as
\be
\vv{R_{r+1}}{L_{r+1}}=
\GG_{r} \PP_r \vv{R_r}{L_r} + a({\bf p})\chi \Ss_{r} \vv{1}{1},
\ee
where
\begin{subequations}
\bea
\GG_r &=& \frac{1}{2n_{r+1}}\(\begin{array}{cc}
n_{r+1}+n_r & n_{r+1}-n_r \\
n_{r+1}-n_r & n_{r+1}+n_r  \end{array}\)  ,
\\
\PP_r &=& \(\begin{array}{cc}
e^{i \delta_r} & 0 \\
0 & e^{-i \delta_r}  \end{array}\) ,
\\
\Ss_r &=& e^{iv_x \omega x_{r+1}}\frac{\chi_{r+1}-\chi_r}{2\chi}
\(\begin{array}{cc}
1+v_x/n_{r+1} & 0 \\
0 & 1-v_x/n_{r+1}  \end{array}\)\label{Sss} .
\eea
\end{subequations}
These equations have the same form as those derived in \cite{Millar:2016cjp}, but there are three key differences. The first and most important is that the phase of the axion changes between each interface, given by an phase in $\Ss_r$. The second is that the discontinuity in the axion induced $H$-field alters the matrix in $\Ss_r$ from the zero velocity limit, which simply has an identity matrix. This will generally give an overall linear shift of $10^{-3}$. Lastly, the velocity also influences the axion frequency $\omega=m_a(1+v^2/2)$. This primarily affects the propagation of the photon, slightly modifying $\PP_r$. Unlike the other two changes, the transverse velocities of the axion also enter $\PP_r$, rather than just $v_x$.
  
For multiple layers we iterate this equation to derive a relation between the amplitude of the $R,L$ waves in the external media, 0 and $m$:
\bea
\vv{R}{L}_m
&=& \GG_{m-1} \PP_{m-1} \vv{R}{L}_{m-1} + a({\bf p})\chi \Ss_{m-1}\vv{1}{1} \nonumber  \\
&=& \GG_{m-1} \PP_{m-1} \[\GG_{m-2} \PP_{m-2} \vv{R}{L}_{m-2} + a({\bf p})\chi \Ss_{m-2}\vv{1}{1}\]+ a({\bf p})\chi\Ss_{m-1}\vv{1}{1} \nonumber  \\
&=& ... \nonumber \\
 &=& \TT \vv{R}{L}_{0} + a({\bf p})\chi \MM \vv{1}{1}\label{transfereq} 
\eea
where $\TT=\TT_0^m$, with $\TT^a_b$ the transfer matrix from surface $b$ to $a$
\be
\TT^a_b = \GG_{a-1}\PP_{a-1}\GG_{a-2}\PP_{a-2}\GG_{a-3}\PP_{a-3} \, ...\, \GG_{b}\PP_{b} .
\ee
In this notation, $\TT^a_a=\PP_0=\mathbb{1}$. Similarly, we have defined 
\be 
\MM\equiv\sum_{s=1}^m  \TT^{m}_s \Ss_{s-1}.\label{eq:mdef}
\ee

As in \cite{Millar:2016cjp}, we have separated \eqref{transfereq} into two parts. The first part encodes the normal propagation of EM waves through dielectric media. The second gives the axion-DM induced source terms $\Ss_r$, one for each layer. Because each wave has to traverse different layers to reach the detector they will all come with different phases. These are the phases we shall arrange to be coherent to increase the axion DM signal. Interestingly, the main axion velocity effects only enter in $\Ss_r$, and then only by $v_x$. Essentially, the dielectric haloscope picks out the part of the axion's velocity aligned with the haloscope. Because of this, in situations where the velocity is non-negligible a dielectric haloscope should have a directional sensitivity to the axion. 
\subsubsection*{Transmissivity and Reflectivity }
We can use $\TT$ to define the usual transmission and reflection amplitude coefficients
\begin{subequations}
\label{eq:T-R-Coefficients}
\begin{eqnarray}
{\cal T}_L &=& \frac{R_m}{R_0}\Big{|}_{L_m=0} = \frac{{\rm Det} [\TT]}{\TT[2,2]}\,, \\[1ex]
{\cal T}_R &=& \frac{L_0}{L_m}\Big{|}_{R_0=0}  =  \frac{1}{\TT[2,2]}\,,  \\[1ex]
{\cal R}_L &=&  \frac{L_0}{R_0}\Big{|}_{L_m=0}  = -\frac{\TT[2,1]}{\TT[2,2]}\,, \\[1ex]
{\cal R}_R &=&  \frac{R_m}{L_m}\Big{|}_{R_0=0} = \frac{\TT[1,2]}{\TT[2,2]}\,,
\end{eqnarray}
\end{subequations}
with Det$[\TT]={n_0}/{n_m}$. These expressions correspond to decoupling the axion by taking $\chi\to 0$, though for $\chi> 0$ the RHS of the above equations \eqref{eq:T-R-Coefficients} then define the transmissivity and reflectivity. These can be useful quantities, as they will generally be correlated with the production of photons from axions. Note that non-zero axion velocities can make this correlation less manifest, as the transmissivity and reflectivity are unaffected by the axion. To regain the connection, one must also know the axion velocity.
\subsubsection*{Boost amplitude}
For a dielectric haloscope we only want the EM waves generated in the presence of axions. In this case we exclude any incoming EM waves, $R_0=L_m=0$, and solve for the outgoing waves to define the usual boost amplitude for dielectric haloscopes
\begin{subequations}
\bea
{\cal B}_L= \frac{L_0}{\chi a({\bf p})}  &=& -\frac{\MM[2,1]+\MM[2,2]}{\TT[2,2]},   \\
{\cal B}_R = \frac{R_m}{\chi a({\bf p})}&=& \MM[1,1]+\MM[1,2] -\frac{\MM[2,1]+\MM[2,2]}{\TT[2,2]}\TT[1,2].
\eea
\end{subequations}
In general, we will not need to worry about the distinction between ${\cal B}_L$ and ${\cal B}_R$, as we only consider situations that are either symmetric, so ${\cal B}_L={\cal B}_R\equiv {\cal B}$, or maximally asymmetric, with the medium on one side being a mirror. In this case only one of ${\cal B}_L$ and ${\cal B}_R$ is non-zero.
To measure the gain in the amplitude of the emitted $E$-field, we define the boost factor $\beta\equiv |{\cal B}|$. Note that usually the detector will be outside the magnetic field, so $E_a$ is not directly measured. Thus the power per unit area emitted in EM waves is
\be
\frac{P({\bf p})}{A}=\frac{E({\bf p})^2}{2}=\chi^2 \omega^2\beta^2({\bf p})|a({\bf p})|^2.\label{eq:power}
\ee
Note that any measurement implicitly happens over some given frequency range (or alternatively, integrating over the corresponding momentum distribution).

\section{Velocity effects in dielectric haloscopes}
\label{analytic}
We can now turn our attention to the effects on the boost factor of dielectric haloscopes caused by the non-zero velocity of the axion. Here we will neglect the finite transverse size of realistic dielectric disks, as well as any potential tilting effects coming from nonparallel disks, both of which may lead to losses. To see the general behaviour of our systems, we revisit the analytic examples studied in~\cite{Millar:2016cjp}. We will start with some general considerations based on the structure of $\MM$ in section~\ref{generalM}. Then we will study a single dielectric disk to show the small shift due to the axion induced $H$-field in section~\ref{singledisk}. We will then see that resonance conditions do not lead to additional velocity effects by studying a simple cavity setup in section~\ref{cavity}. We will then see from a transparent setup containing many disks that in section~\ref{transparent} the primary effect comes from the linear extent of the dielectric haloscope in the $x$ direction, i.e., the number of disks, their spacings and to a lesser extent the thickness of the disks. Velocity effects start having a ${\cal O}(10\%)$ effect when the length of haloscope in the $x$ direction ($L$) is $15-20$\% of the de Broglie wavelength of the axion. When the de Broglie wavelength is similar to the length of the haloscope the axion velocity can give an ${\cal O}(1)$ effect. To show that this remains true for non-trivial configurations of the type that would be used by an experiment, we explore the effect of a non-zero velocity on a realistic 80 disk setup in section \ref{80disk}.

\subsection{General behaviour}
\label{generalM}
Before we begin studying explicit examples, it would be instructive to gain a general understanding of at what order and from which terms velocity effects begin to affect dielectric haloscopes. To do this, we must look closer at $\MM$ (as defined in \eqref{eq:mdef}). We note that it is possible to rewrite any element of a transfer matrix $\TT_a^b[i,j]$ in the form
\be
\TT_a^b[i,j]=\sum_{p} W^{i,j}_{p}e^{i\delta_{p}},
\ee
where $W_p^{i,j}$ is a constant depending only on the $n_r$, and $\delta_{p}$ is a combination of the phases with differing signs, i.e., each $\delta_p$ is of the form $\pm \delta_a\pm...\pm\delta_b$ with all $2^{b-a-1}$ possible combinations.
From this we can write
\bea
\MM[i,j]=\sum_{s=1}^m e^{iv_x \omega x_{s}}\frac{\chi_{s}-\chi_{s-1}}{2\chi}
\(1\pm\frac{v_x}{n_{s}}\)\sum_{p}W^{i,j}_{p}e^{i\delta_p},
\eea
where the $\pm$ signs depends on the exact matrix element. The axion velocity enters in three places: the change in the phase of the axion, the discontinuity in the axion induced $H$-field and the shift to the axion frequency.
 To isolate the velocity effects, we can look at
\bea
v_x\frac{\partial \MM[i,j]}{\partial v_x}&\supset & v_x\sum_{s=1}^mi\omega x_s\frac{\chi_{s}-\chi_{s-1}}{2\chi}\sum_{p} W^{i,j}_{p}e^{i\delta_p} \big{|}_{v=0}\nonumber \\
&\pm &v_x\sum_{s=1}^m\frac{1}{n_s}\frac{\chi_{s}-\chi_{s-1}}{2\chi}\sum_p W^{i,j}_pe^{i\delta_p}
 \big{|}_{v=0}\, \nonumber \\
&+&
 v_x^2\sum_{s=1}^m\frac{\chi_{s}-\chi_{s-1}}{2\chi}\sum_p i\delta_p W^{i,j}_{p}e^{i\delta_p} \big{|}_{v=0}\, ,\label{eq:genv}
\eea
where we have only kept the leading contribution from each of the three effects mentioned above. This assumes that the phase appearing in the exponentials is small, which breaks down for physically large haloscopes. The three effects can be understood separately: 
\begin{enumerate}
\item{{\em Axion phase:} The term in the first line of \eqref{eq:genv} is the most important, and comes from the change of phase of the axion itself. As it comes with factors of $x_s$, this term is roughly proportional to $m_a L v_x$, where $L$ is the length of the haloscope in the $x$ direction, leading it to dominate over the other terms as $L$ increases. When $L$ is very large, this expansion breaks down and the change in $\MM$ is significant, potentially even ${\cal O}(1)$. Note that this is only the leading contribution coming from the phase of the axion: we neglect cross terms that may be of the same or higher order as the other two contributions, such as terms proportional to $m_a^2 L^2 v_x^2$ and $m_a Lv_x^2$.}

\item{{\em $H$-field:}  The term in the second line of \eqref{eq:genv} is due to the discontinuity in the axion induced $H$-field. Note that there is no enhancement by the length scales of the haloscope. Because of this, terms coming purely from the axion induced $H$-field cannot have a significant effect on $\MM$ as $v$ is very small.}

\item{{\em Frequency shift:} The contribution in the last line of \eqref{eq:genv} comes from the frequency shift of the axion appearing in the photon propagator. We can see from the combination of the phase $\delta_p$ and the factor of $v_x^2$ that this contribution is generically proportional $m_a Lv^2$. As this is a frequency shift, it can only large effect when the boost factor varies rapidly as a function of frequency, a case which is not usually considered in the dielectric haloscope context. For the cases we will study below, $\beta(m_a)\simeq\beta(\omega)$ to very good approximation.
Note that the transverse velocities also result in a similar frequency shift. For simplicity, we will assume throughout this section that the disk is perfectly aligned with the axion velocity, so that $v\equiv v_x$. As the shift to the frequency makes certain analytic cases more complex, but is numerically irrelevant, for the analytic cases we will define distances in terms of $\omega$, rather than the axion mass~$m_a$. Note that even though the change to the boost factor is negligible, a detector with an energy resolution better than ${\cal O }(10~{\rm kHz})$ would be able to resolve this frequency shift for $100~\mu$eV axions. }
\end{enumerate}

Thus we expect that the primary influence on the boost factor will come from the change of phase of the axion. This can be contrasted with the velocity effects considered in dish antenna setups~\cite{Jaeckel:2013sqa,Jaeckel:2015kea}. Dish antennas have only a single interface, essentially giving $L=0$ so that the dominant effects come from the axion velocities parallel to the interface. 
To see exactly at what level the velocity comes into play, we must consider some more concrete examples.

\subsection{Single dielectric disk}
\label{singledisk}
As they form the building blocks of the dielectric haloscope, it is instructive to consider a single dielectric disk, as depicted in figure \ref{fig:disk}. This disk can be categorised by a thickness $d$ and an index of refraction $n\equiv n_1>1$, with vacuum on either side ($n_0=n_2=1$). The change of phase experienced by a perpendicular EM wave inside this disk is thus $\delta\equiv \delta_1=n\omega d$. Using transfer matrices, equation~\eqref{transfereq}, we can then solve for 
\begin{subequations}
\bea
{\cal T}_{\rm D}&=&\frac{i\,2n}{i\,2n\cos\delta+(n^2+1)\,\sin\delta}\, ,\\
{\mathcal B}_{\rm D}&=&\frac{(n^2-1)\[ \(e^{2i\delta}-1\)n^2+n(v-1)\(1+e^{2i\delta}-e^{\frac{i\delta(n+v)}{n}}\)+v+ ve^{2i\delta}\]}{n^2[(n-1)^2e^{2i\delta}-(n+1)^2]},
\eea
\end{subequations}
where we have used the subscript D to indicate that these quantities are for a single disk. These expressions agree with the corresponding ones in \cite{Millar:2016cjp} when $v=0$. Note that ${\cal B}_{\rm D}$ depends on $v$, whereas ${\cal T}_{\rm D}$ does not. While a strong correlation between ${\cal B}_{\rm D}$ and ${\cal T}_{\rm D}$ was found in \cite{Millar:2016cjp}, if velocity effects are significant this correlation is less manifest. To gain information about the boost factor from the transmissivity, one must also know the axion velocity.
\begin{figure}[t]
\begin{center}
\includegraphics[width=6cm]{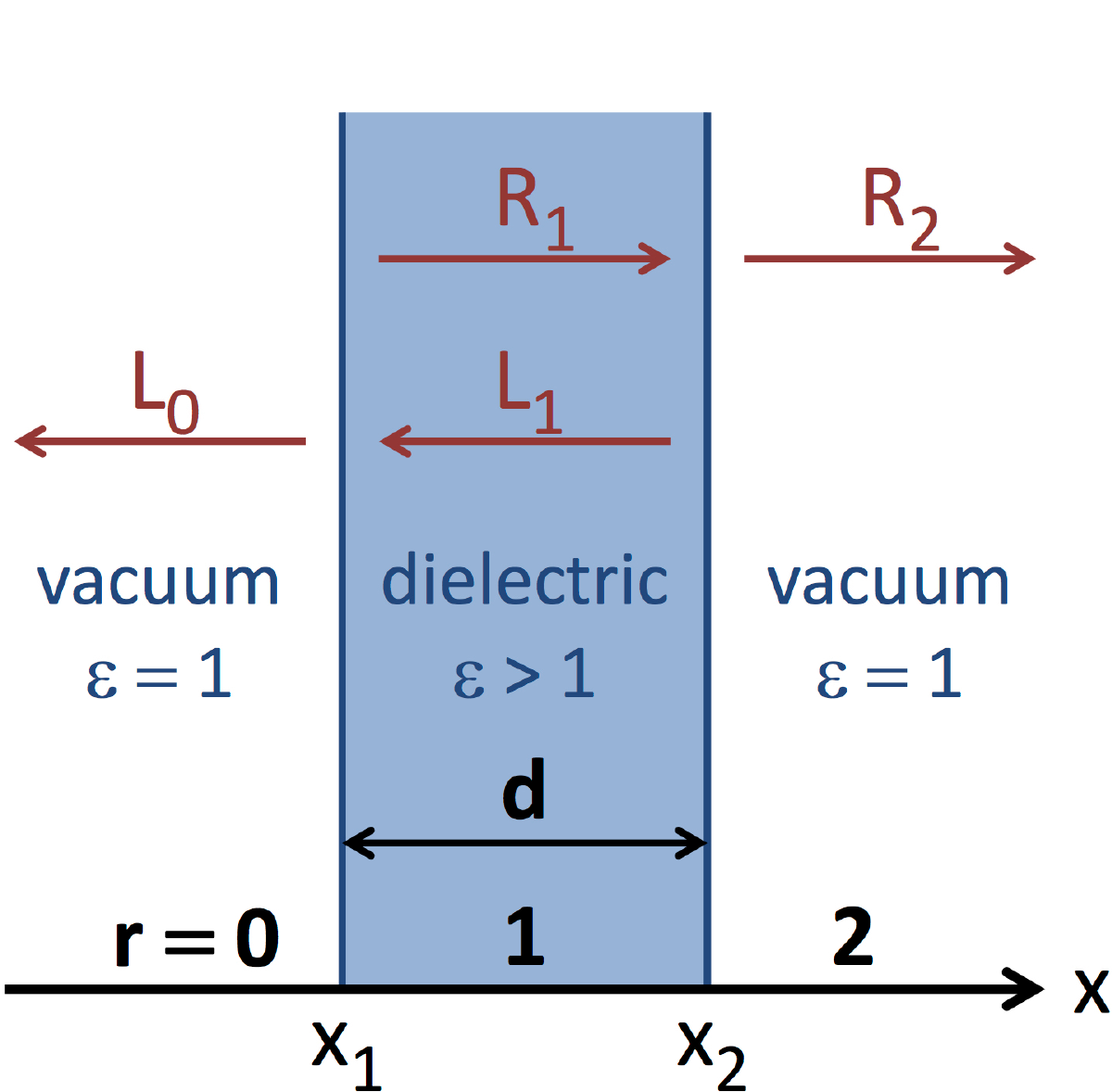}
\caption{Schematic of a single dielectric disk with dielectric constant $\epsilon$ and disk thickness $d$.}
\label{fig:disk}
\end{center}
\end{figure}

For a simple example, we can consider a transparent disk, which is characterised by $\delta=\pi$.
Expanding to linear order in $v$ gives
\bea
{\cal B}_{\rm D}&=&\frac{1-v}{2}\(1-\frac{1}{n^2}\)\(1+e^{\frac{i\pi v}{n}}\)\nonumber \\
&\sim & \(1-\frac{1}{n^2}\)\(1-v+\frac{i\pi v}{2n}\).
\eea
In the second line of this equation, the term proportional to $-v$ can be understood as coming from the direct modification to $\Ss_r$, i.e., the change to $H_a$. The second velocity dependent term, proportional to $\frac{i\pi v}{2n}$, comes from the change in phase the axion experiences over the disk, as seen from the factor of $\pi/2$. However, as this second term is the only contribution to $\rm {Im}({\cal B}_{\rm D})$, it will be relegated to second order in $\beta_{\rm D}=|{\cal B}_{\rm D}|$, leaving only the term coming from the axion induced $H$-field
\be
\beta_{\rm D}\sim(1-v)\(1-\frac{1}{n^2}\).
\ee
As one would expect, for a single dielectric disk with a reasonably small thickness $d$ the velocity of the axion has only a tiny effect on $\beta_{\rm D}$. 

As they form the building blocks of the dielectric haloscope, it is instructive to consider a single dielectric disk, as depicted in figure \ref{fig:disk}. This disk can be categorised by a thickness $d$ and an index of refraction $n\equiv n_1>1$, with vacuum on either side ($n_0=n_2=1$). The change of phase experienced by a perpendicular EM wave inside this disk is thus $\delta\equiv \delta_1=n\omega d$. Using transfer matrices, equation~\eqref{transfereq}, we can then solve for 
\begin{subequations}
\bea
{\cal T}_{\rm D}&=&\frac{i\,2n}{i\,2n\cos\delta+(n^2+1)\,\sin\delta}\, ,\\
{\mathcal B}_{\rm D}&=&\frac{(n^2-1)\[ \(e^{2i\delta}-1\)n^2+n(v-1)\(1+e^{2i\delta}-e^{\frac{i\delta(n+v)}{n}}\)+v+ ve^{2i\delta}\]}{n^2[(n-1)^2e^{2i\delta}-(n+1)^2]},
\eea
\end{subequations}
where we have used the subscript D to indicate that these quantities are for a single disk. These expressions agree with the corresponding ones in \cite{Millar:2016cjp} when $v=0$. Note that ${\cal B}_{\rm D}$ depends on $v$, whereas ${\cal T}_{\rm D}$ does not. While a strong correlation between ${\cal B}_{\rm D}$ and ${\cal T}_{\rm D}$ was found in \cite{Millar:2016cjp}, if velocity effects are significant this correlation is less manifest. To gain information about the boost factor from the transmissivity, one must also know the axion velocity.
\begin{figure}[t]
\begin{center}
\includegraphics[width=6cm]{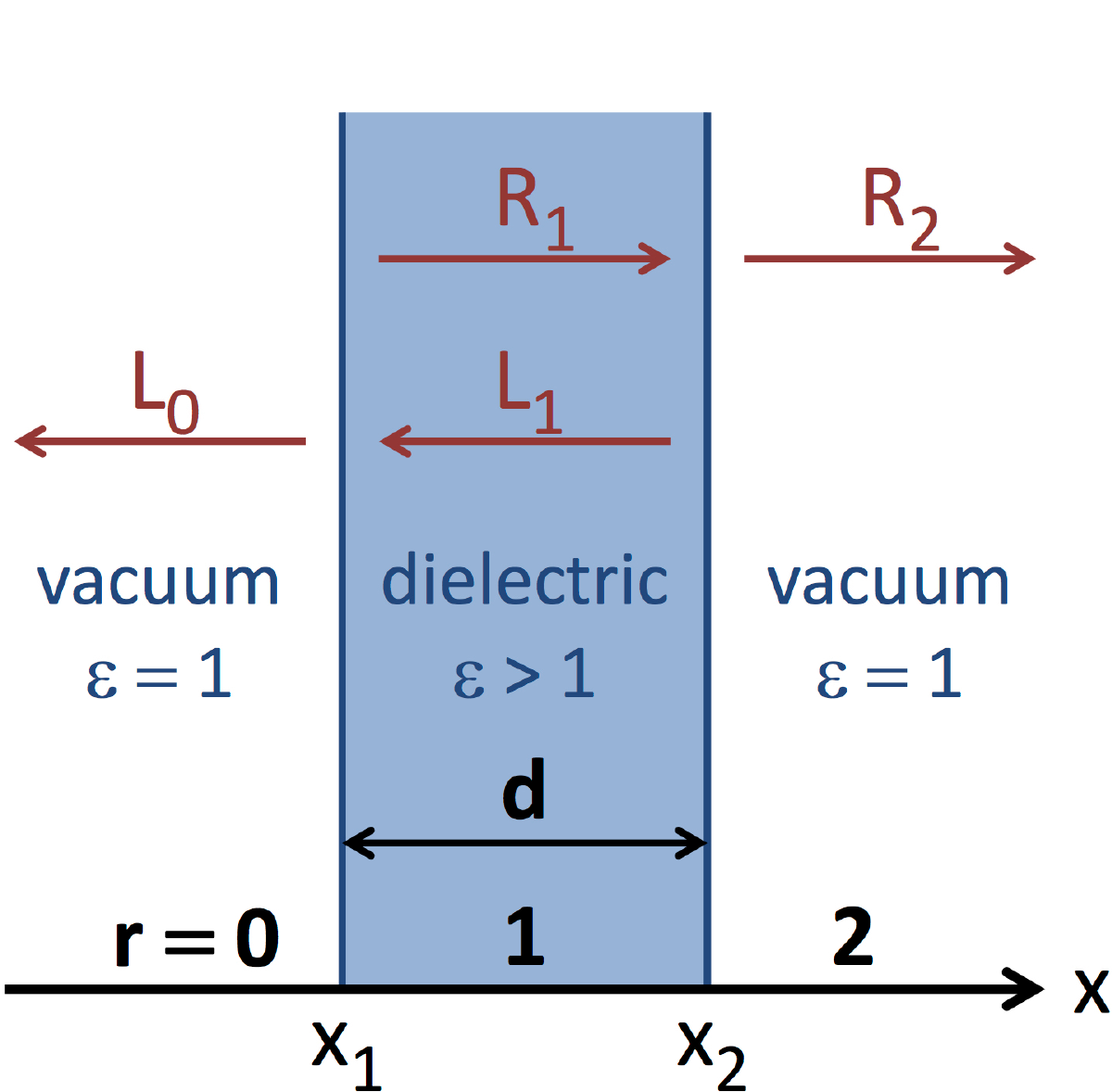}
\caption{Schematic of a single dielectric disk with dielectric constant $\epsilon$ and disk thickness $d$.}
\label{fig:disk}
\end{center}
\end{figure}

For a simple example, we can consider a transparent disk, which is characterised by $\delta=\pi$.
Expanding to linear order in $v$ gives
\bea
{\cal B}_{\rm D}&=&\frac{1-v}{2}\(1-\frac{1}{n^2}\)\(1+e^{\frac{i\pi v}{n}}\)\nonumber \\
&\sim & \(1-\frac{1}{n^2}\)\(1-v+\frac{i\pi v}{2n}\).
\eea
In the second line of this equation, the term proportional to $-v$ can be understood as coming from the direct modification to $\Ss_r$, i.e., the change to $H_a$. The second velocity dependent term, proportional to $\frac{i\pi v}{2n}$, comes from the change in phase the axion experiences over the disk, as seen from the factor of $\pi/2$. However, as this second term is the only contribution to $\rm {Im}({\cal B}_{\rm D})$, it will be relegated to second order in $\beta_{\rm D}=|{\cal B}_{\rm D}|$, leaving only the term coming from the axion induced $H$-field
\be
\beta_{\rm D}\sim(1-v)\(1-\frac{1}{n^2}\).
\ee
As one would expect, for a single dielectric disk with a reasonably small thickness $d$ the velocity of the axion has only a tiny effect on $\beta_{\rm D}$.

\subsection{Cavity mode}
\label{cavity}
\begin{figure}[t]
\begin{center}
\includegraphics[width=8cm]{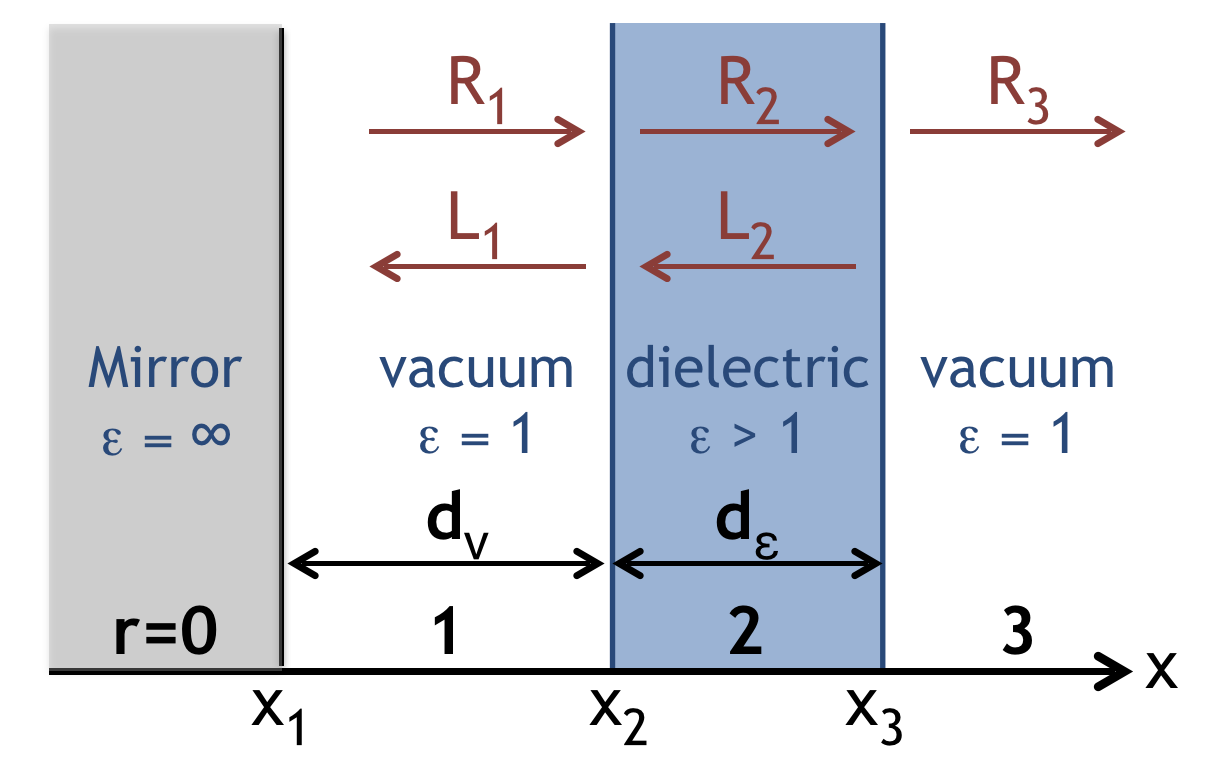}
\caption{Schematic of a simple cavity consisting of a mirror and a single dielectric disk. The distances $d_{\rm v}$ and $d_\epsilon$ denote the thicknesses of the vacuum gap and dielectric disk, respectively. Note that $R_0,L_0$ (not shown) are both zero inside the mirror.}
\label{fig:cavity}
\end{center}
\end{figure}

One might wonder whether velocity effects can be enhanced by multiple reflections forming a resonance. To check this, we can take the simple cavity setup from section 5.2 of \cite{Millar:2016cjp}, consisting of a single mirror with a dielectric disk of refractive index $n$ (shown in figure~\ref{fig:cavity}). The phase change in the vacuum gap $\delta_{\rm v}$ and the disk $\delta_\epsilon$ can be chosen so that a resonance occurs: $\delta_{\rm v}=(2m-1)\pi$, where $m\in{\mathds N}$, and $\delta_\epsilon=\pi/2$. Such a resonance can be handled analytically, giving 
\be
{\cal B}_{\rm C}=n\(e^{\frac{i\pi v}{2n}}+e^{\frac{i\pi v[ 1+2n(2m-1)]}{2n}}\)-\frac{ie^{\frac{i\pi v}{2n}}}{n}-v\(1-\frac{1}{n^2}\), \label{eq:bcav}
\ee
where the C subscript stands for cavity.
There are two instructive limits of this equation: a strongly resonant setup and a physically large setup (i.e., large $m$). 

Let us first check the strongly resonant limit of this equation. The resonance will give a sharp, high boost factor peak when the disk is as reflective as possible, i.e., when $n$ is large. Expanding \eqref{eq:bcav}, taking the lowest order terms in $mv$ and assuming that $n\gg 1$ we get  
\be
\beta_C=|{\cal B}_{\rm C}|\sim n\[2-\(\frac{v \pi [2m-1]}{2}\)^2\].
\ee
We can see from this that enhancing the resonant conditions (by increasing $n$) does not change the proportional importance of velocity effects. Because of this we do not expect that resonant setups will be more prone to velocity effects than non-resonant setups, such as those discussed below. Note that here we have assumed that the resonance width is much larger than the axion line width, as is expected for dielectric haloscopes for practical reasons~\cite{Millar:2016cjp}.

On the other hand, if the device is made large by increasing the spacing between the disk and the mirror, the velocity dependent terms can dominate $\beta_{\rm C}$. This can be seen immediately as the dependence on the size of the device (given by $m$) enters as the phase in a complex exponential in \eqref{eq:bcav}. Such a setup was proposed in~\cite{Irastorza:2012jq} to gain a directional sensitivity to the axion velocity. While in principle the axion velocity could enter at linear order, this does not occur due to a $\pi/2$ phase shift between the linear order $v$ terms and the $v$ independent terms in ${\cal B}_{\rm C}$. 

\begin{figure}[b]
\begin{center}
\includegraphics[width=9.5cm]{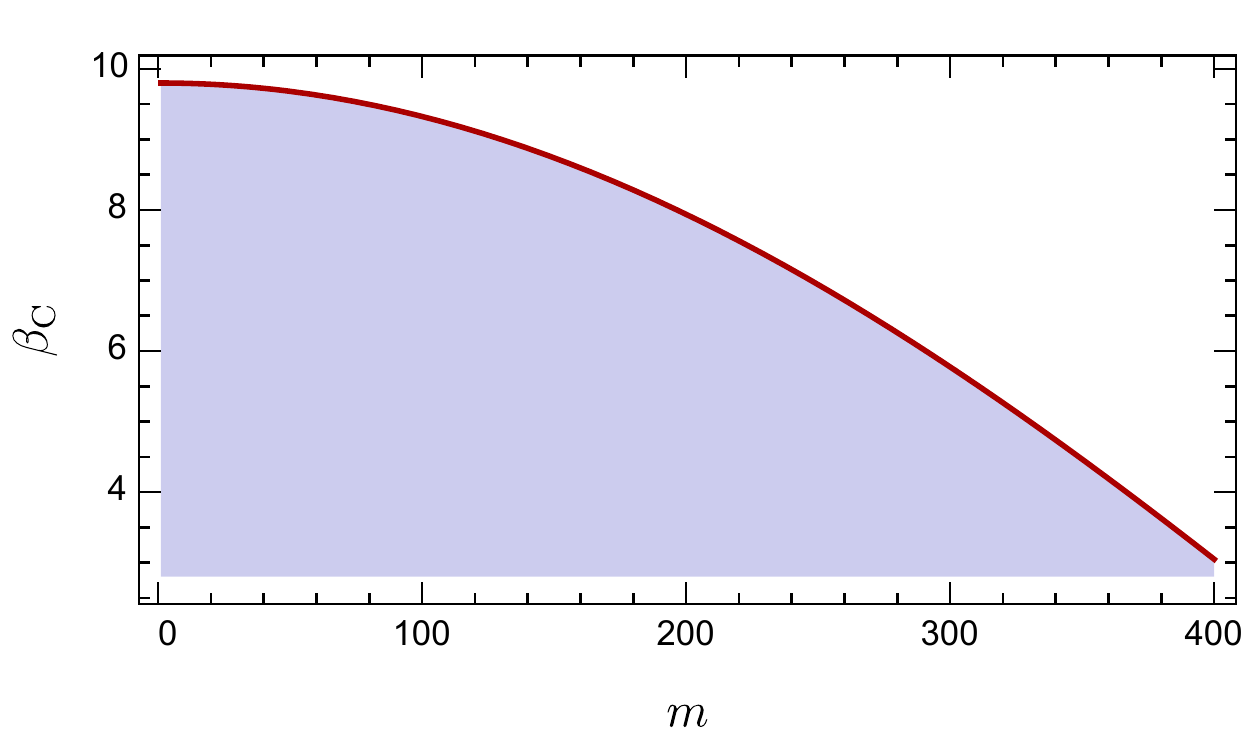}
\caption{$\beta_{\rm C}$ as a function of $m$ for a resonant cavity consisting of a mirror and a single dielectric disk of optical thickness $\pi/2$ and refractive index $n=5$, with $2m-1$ half wavelengths between the mirror and reflective disk. Here we have used $v=10^{-3}$; if $v=0$ then $\beta_{\rm C}$ would be constant as we assume lossless materials. }
\label{fig:cbeta}
\end{center}
\end{figure}

If the phase change of the axion is an appreciable fraction of $2\pi$, i.e., the device is a large fraction of the de Broglie wavelength, velocity effects will be significant. In figure~\ref{fig:cbeta} we plot the boost factor against $m$, with the number of half wavelengths in the device given by $2m-1$. From figure~\ref{fig:cbeta} we see that these effects reach the $10\%$ level at a scale that roughly corresponds to 15\% of the de Broglie wavelength~\eqref{eq:lambdadeBroglie}. Here we have assumed that $n=5$, however this result does not depend strongly on the exact choice of $n$.
Of course, this is not a particularly practical limit: one would never actually choose to place the disk hundreds of wavelengths away from the mirror. The losses caused by diffraction, and the narrowing of the boost factor bandwidth that accompanies higher order modes would make this a very inefficient experiment. However, one could ameliorate the losses due to diffraction by enclosing the experiment in a metallic cavity.

\subsection{Transparent mode}
\label{transparent}
\begin{figure}[t]
\begin{center}
\includegraphics[width=13cm]{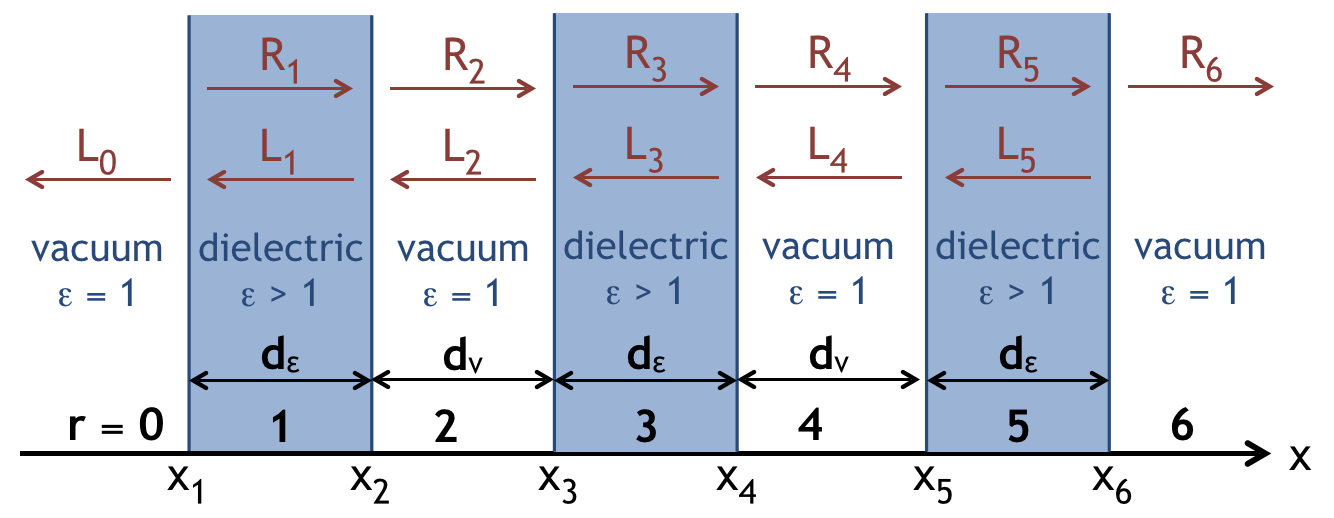}
\caption{Schematic of a simple dielectric haloscope consisting of three dielectric disks, each with refractive index $n=\sqrt {\epsilon}$. The distances $d_{\rm v}$ and $d_\epsilon$ denote the thicknesses of the vacuum gaps and dielectric disks, respectively. }
\label{fig:transparent}
\end{center}
\end{figure}

\begin{figure}[t!]
\begin{center}
\includegraphics[width=9.5cm]{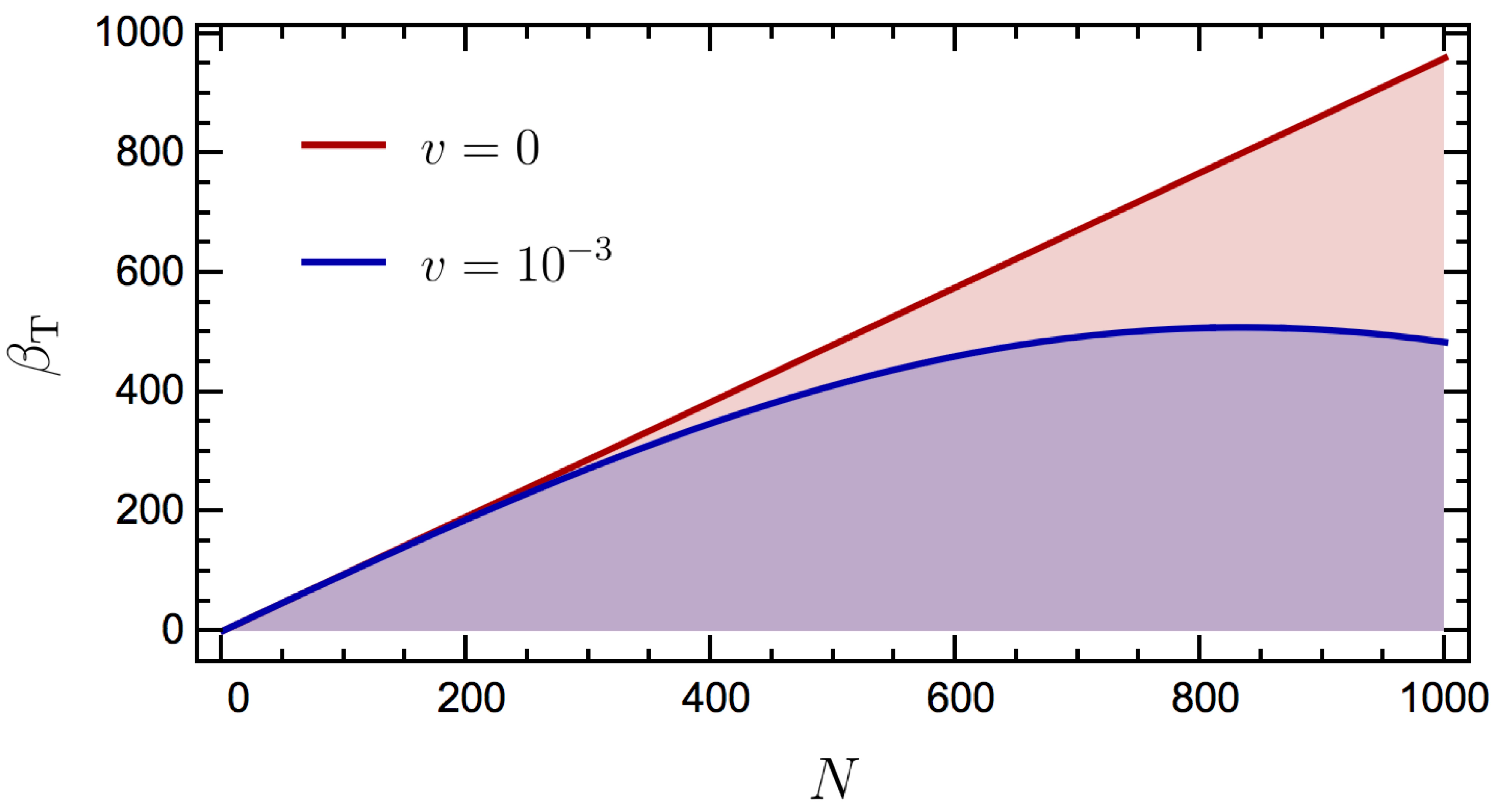}
\caption{Boost factor $\beta_{\rm T}(N)$ for a setup consisting of $N$ transparent (optical thickness $\pi$) dielectric disks of refractive index $n=5$ equally spaced with an optical thickness $\pi$ between each disk. The blue curve corresponds to $v=10^{-3}$, the red to $v=0$.}
\label{fig:tbeta}
\end{center}
\end{figure}

Let us consider next a simple multidisk setup, that consists of a series of transparent disks, each with a refractive index $n$. For some frequency $\omega_0$, we can specify that each disk has an optical thickness $\delta_\epsilon=n\omega_0 d_\epsilon=\pi$, each equally spaced with a vacuum gap of $\delta_{\rm v}=\omega_0 d_{\rm v}=\pi$. A three disk example is shown in figure~\ref{fig:transparent}.
As we have presented the building blocks of the transparent mode in section \ref{singledisk}, we can see how one puts them together. We must be more careful than in the zero velocity limit explored in~\cite{Millar:2016cjp}: the boost factor for $N$ disks is no longer simply $N$ times the boost factor of a single disk. The result still is given by a sum of boost factors but we must also account for the change in the phase of the axion at each interface. Note that both the right and the left side of each interface must be summed over, giving two sums
\bea
{\cal B}_{\rm T}&=&\(1-v\)\(1-\frac{1}{n^2}\)\[\sum_{j=1}^{N}e^{\frac{i\pi v j}{n}+(j-1)i\pi v}\nonumber -\sum_{j=1}^{N-1}e^{\frac{i\pi v j}{n}+i\pi v j}\] \\
&=&N\frac{1-v}{2}\(1-\frac{1}{n^2}\)\frac{\(1+e^{\frac{i\pi v}{n}}\)\(e^{\frac{i(1+n)N\pi v}{n}}-1\)}{e^{\frac{i(1+n)\pi v}{n}}-1}.
\eea
We have used the subscript $\rm T$ to denote that this boost amplitude is for the transparent mode. Similar to the cavity case, as velocity effects come in from a phase enhanced now by a factor of $N$, for large numbers of disks, corresponding to a physically large device, they will govern ${\cal B}_{\rm T}$. As $N$ can be arbitrarily high in principle, to see the full behaviour we must in principal keep all orders of $Nv$. However, a series expansion can help us see at what $N$ velocity effects become important. The most relevant terms at large $N$ are
\be
\beta_{\rm T}/N\sim 1-\frac{N^2\pi^2v^2}{12}\, .\label{eq:betaT2}
\ee
From this it is clear that when $N$ is large the velocity dependent terms will dominate, entering at quadratic order. Again, there are no linear order terms in $\beta_{\rm T}$ due to a relative phase between the relevant linear order velocity term and the velocity independent terms.
To see when velocity effects become important we plot $\beta_{\rm T}$ in figure \ref{fig:tbeta}, using dielectric disks with $n=5$. While the exact point this occurs at depends weakly on $n$, for $v=10^{-3}$ velocity effects reach the $10\%$ level for $N\gtrsim 400$. This roughly corresponds to a linear distance $L=(N-1)\lambda/2+N\lambda/2n$ of $20\%$ of the axion de Broglie wavelength.

\begin{figure}[t]
\begin{center}
\includegraphics[width=7.6cm]{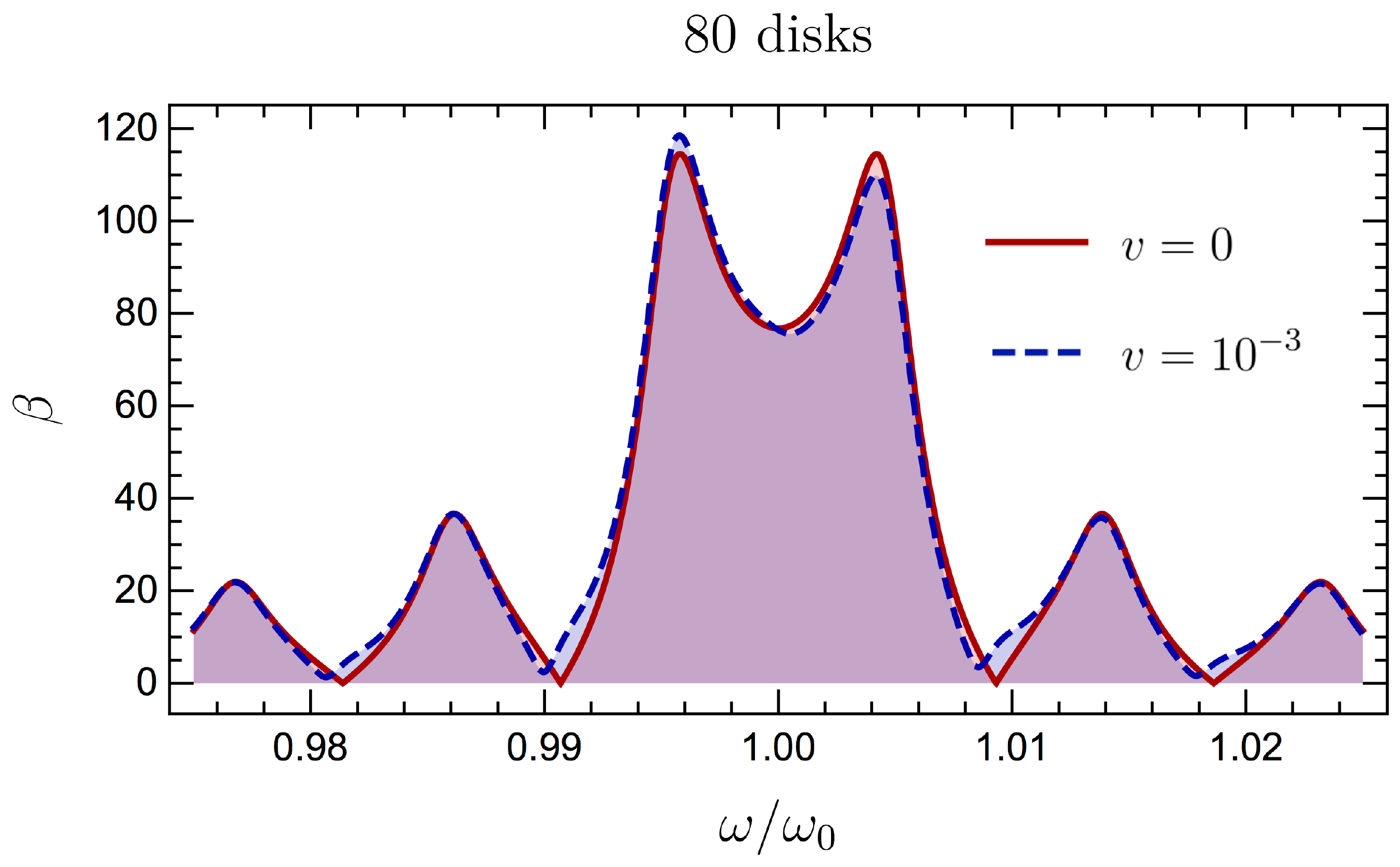}
\includegraphics[width=7.6cm]{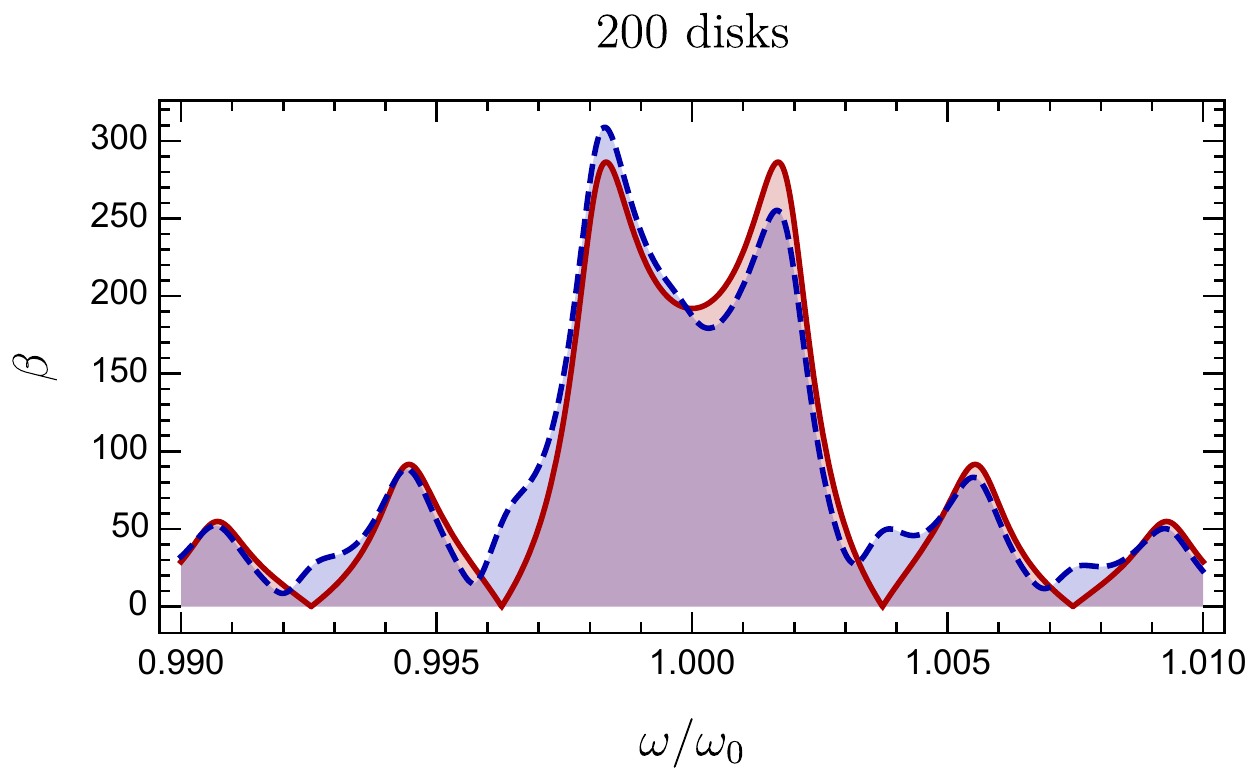}
\includegraphics[width=7.6cm]{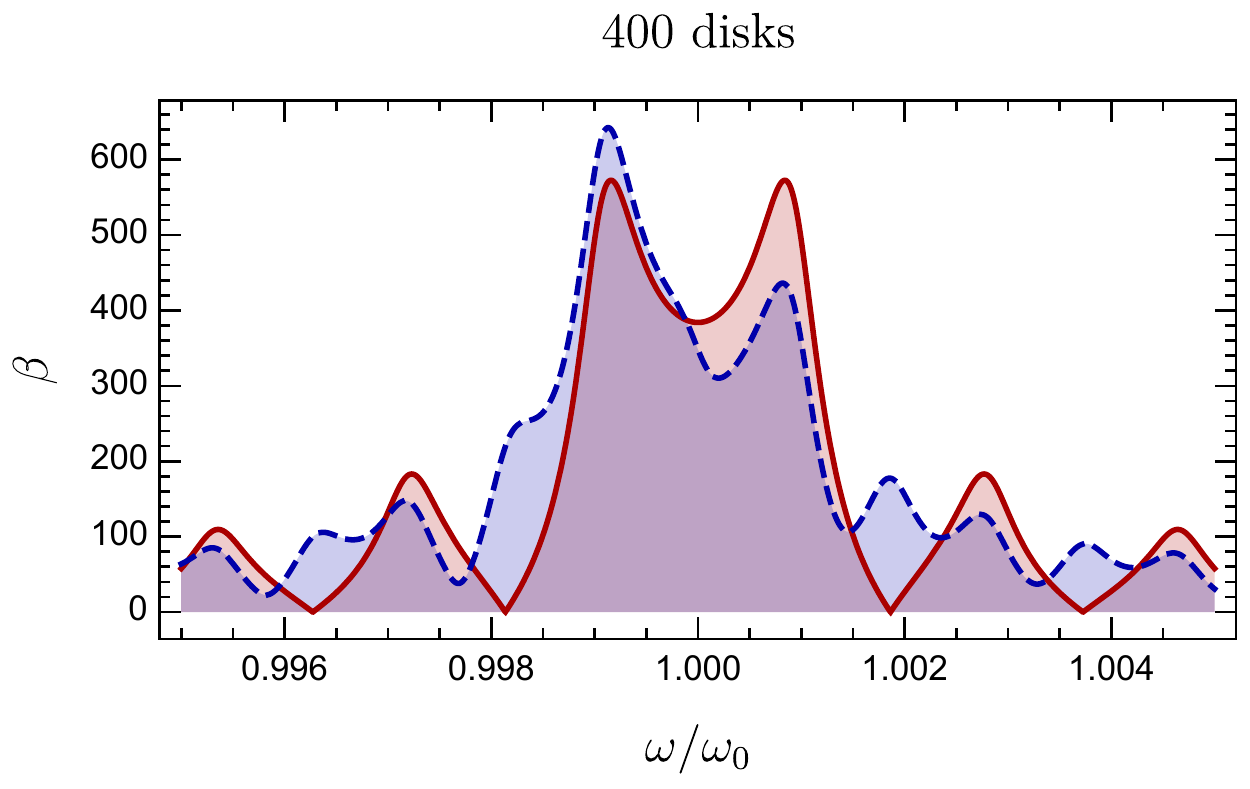}
\includegraphics[width=7.6cm]{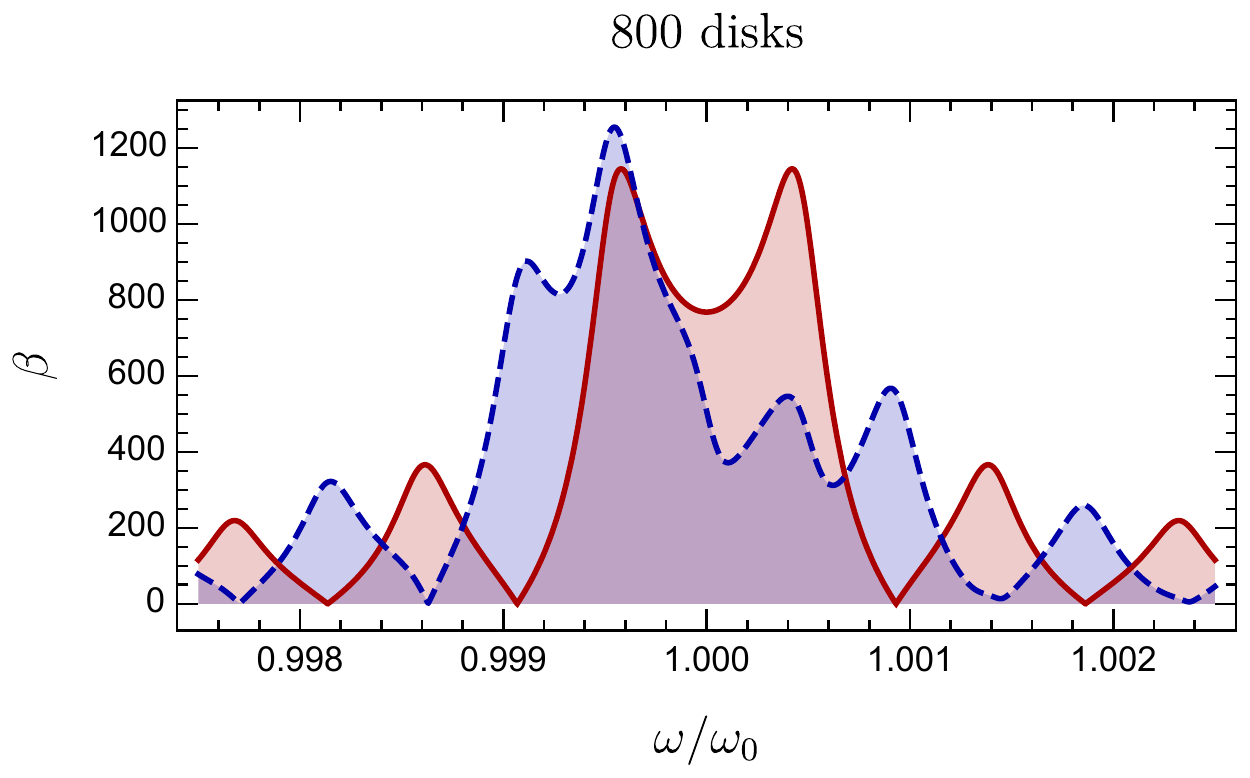}
\caption{Boost factor $\beta(\omega/\omega_0)$ for 80, 200, 400 and 800 disks (top left, top right, bottom left and bottom right panels, respectively). Note that $\omega=\omega_0$ corresponds to the transparent mode, in which all disks are transparent. In each case, the dashed blue curve corresponds to $v=10^{-3}$, the solid red curve to $v=0$. }
\label{fig:Tdisks}
\end{center}
\end{figure}

Note that the transparent mode is an example of a broadband frequency response: in general one is also interested in the behaviour of frequencies adjacent to $\omega_0$.
By looking at the adjacent frequencies, as depicted in figure~\ref{fig:Tdisks}, we see that some frequencies are more prone to velocity effects than others. For 80 disks, as considered in \cite{TheMADMAXWorkingGroup:2016hpc,Millar:2016cjp}, we see that velocity effects are only on the percent level. Going up to 200 disks we start to see an effect, with 400 disks showing ${\cal O}(10\%)$ deviation from the $v=0$ case. At 800 disks there is an ${\cal O}(1)$ change to $\beta$. However, these changes are not uniform: for example for 800 disks there are frequencies that go from $\beta\sim 0$ to $\beta\sim 800$, and others that exhibit almost no change. The differing response to the axion velocity at different frequencies is due to the relative phase structure of the velocity independent and dependent terms. As there is still a high degree of symmetry to the system the velocity effects show a periodic variation of maximal constructive/destructive interference. 

\begin{figure}[t]
\begin{center}
\includegraphics[width=9.5cm]{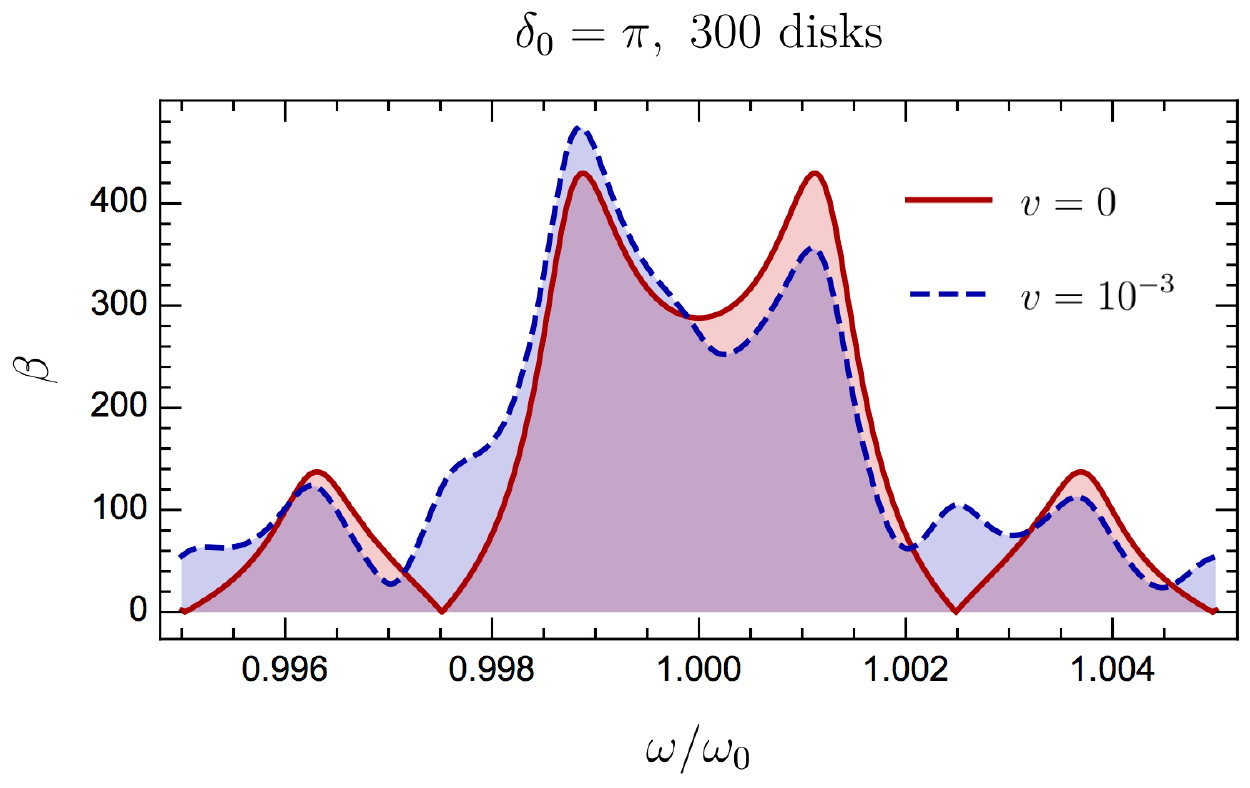}
\includegraphics[width=9.5cm]{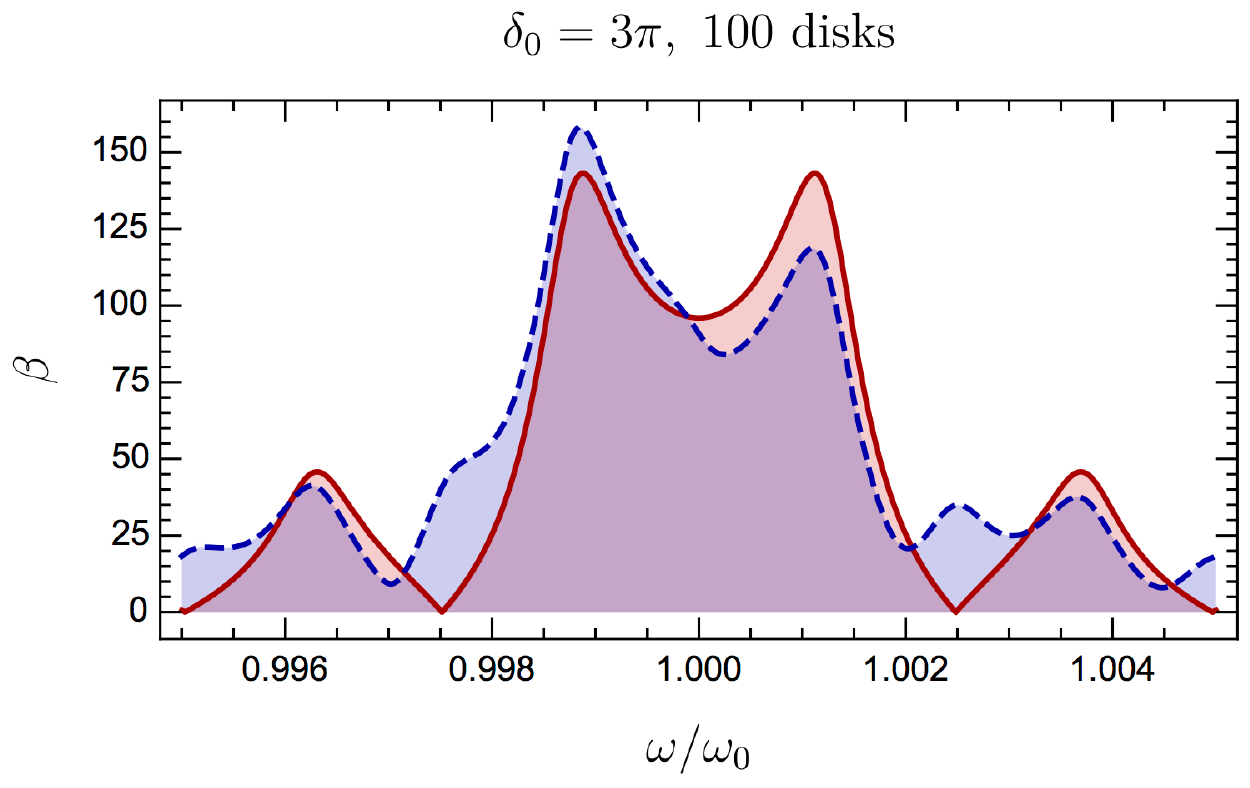}
\caption{Boost factor $\beta(\omega)$ for 300 disks with $\delta_0\equiv\delta_{\rm v}\equiv\delta_\epsilon=\pi$ for $\omega=\omega_0$ (top) and 100 disks with $\delta_0=3\pi$ (bottom). Note that $\omega=\omega_0$ corresponds to the transparent mode, in which all disks are transparent. In each case, the blue curve corresponds to $v=10^{-3}$, the red to $v=0$. }
\label{fig:ddisks}
\end{center}
\end{figure}

While it appears that this effect is coming from adding disks, this is misleading. One can see the same effects by simply increasing the spacing between the disks, as in section~\ref{cavity}. As shown in figure~\ref{fig:ddisks} the same fractional deviation from the zero velocity limit occurs for 100~disks with $3\pi$ phase shifts as for 300 disks with $\pi$ phase shift (though the latter will achieve a higher boost factor as it uses more disks). This is because the exponentiated factors of $N$ come from the change of phase of the axion over the haloscope. Note that increasing the spacing of the disks also results in a narrowing of the boost factor peak, though also allows new structures to exist (i.e., at $\omega=\omega_0/3$). So again we see that the primary influence comes simply from the size of the de Broglie wavelength of the axion when compared to the physical size of the dielectric haloscope.

\subsection{80 disk dielectric haloscope}
\label{80disk}
While the analytically tractable cases discussed above are illustrative, it is also important to check that realistic setups behave similarly. For an experiment it is desirable to have both
broad, rectangular (top-hat) responses and narrow resonant responses \cite{TheMADMAXWorkingGroup:2016hpc}.

To see this, we must consider how an experiment would be performed. The boost factor, while potentially broadband, does not cover the full frequency range we would like to search. Thus we must scan across frequencies. To do so, one must arrange the disks to enhance $\beta$ over a given frequency range, measure, then adjust the positions of the disks to measure an adjacent frequency range and repeat untill the full range is covered. For each measurement taking a time $\Delta t$, Dicke's radiometer equation gives the signal to noise ratio 
\be
\frac{S}{N}=\frac{P}{T_{\rm sys}}\sqrt{\frac{\Delta t}{\Delta\nu_a}},
\ee
where the $T_{\rm sys}$ is the system noise temperature and $\Delta\nu_a\sim 10^{-6}\nu$ is the axion line width. Recall that the power $P$ is given by \eqref{eq:power}. As the Area Law for dielectric haloscopes implies that $P\Delta\nu_\beta$ is roughly constant (where $\Delta\nu_\beta$ is the frequency width over which $\beta$ is enhanced), naively one expects that a narrow resonance with high $P$ is desirable~\cite{Millar:2016cjp}. However, for an optimal scan rate, the measurement time should be similar to the time taken to adjust the disks. Broadband setups can compensate for the potentially lengthy time required to readjust the disk positions between each measurement. 
Such responses would allow one to scan a large frequency range in a single measurement; one could then use narrow resonances to confirm or reject a potential discovery at a higher signal to noise ratio. As in general the axion frequency is not simply related to the disk thickness (such as by a half wavelength as in the transparent mode above), the disk configurations must be found by a numerical optimisation procedure, as described in~\cite{Millar:2016cjp}. In the above examples we saw that the scale at which velocity effects become important depends almost solely on the length of device relative to the de Broglie wavelength~\eqref{eq:lambdadeBroglie}. In particular, we found that resonance effects are unimportant. Thus we expect that our general considerations hold for both broadband and resonant disk configurations.

\begin{figure}[t]
\begin{center}
\includegraphics[width=10cm]{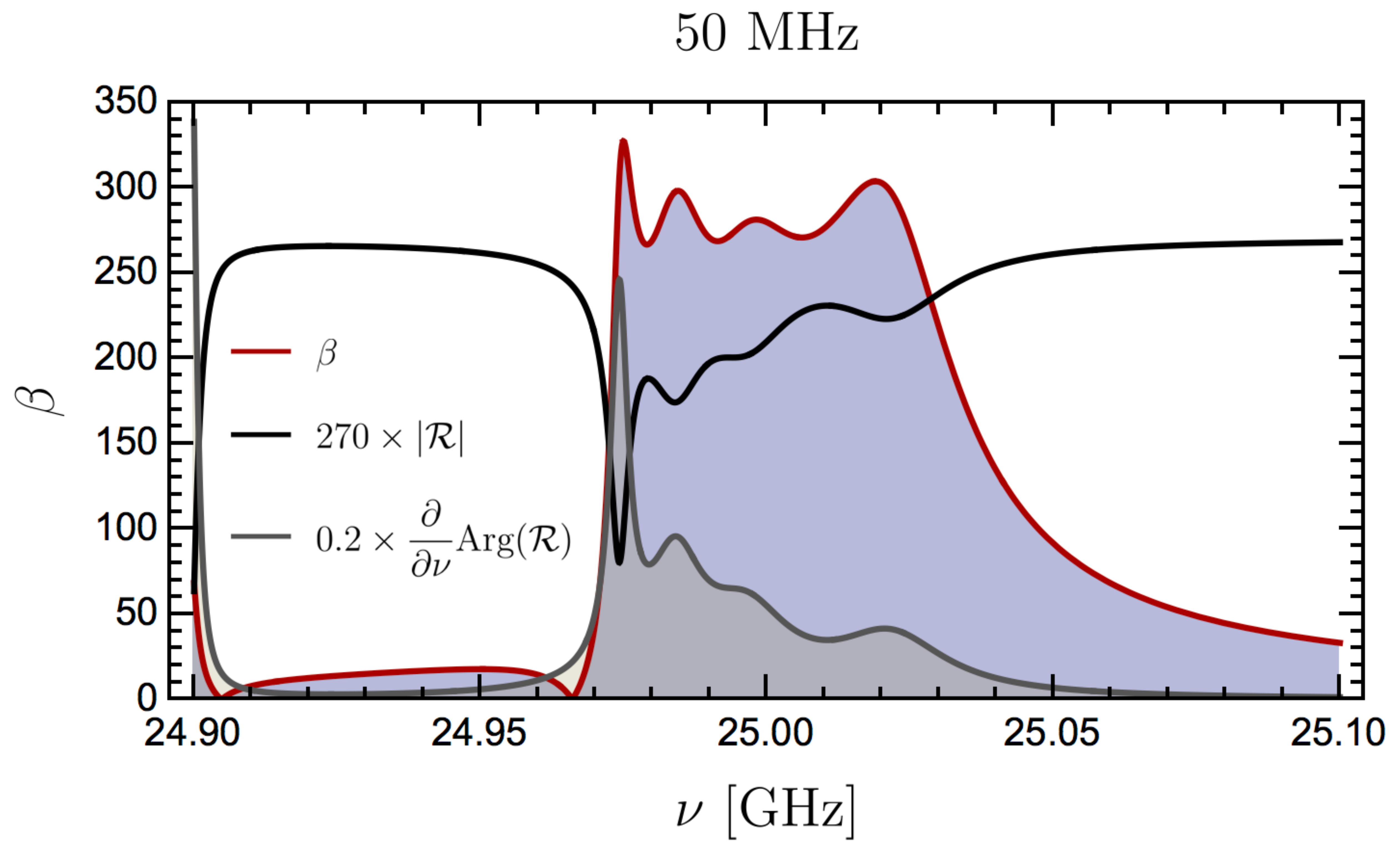}
\caption{Boost factor $\beta(\nu)$ (red curve), reflectivity $|{\cal R}(\nu)| $ (black curve) and group delay $\frac{\partial}{\partial \nu}\text {Arg}(\mathcal R)$ (gray curve) for 80 disks optimised for a bandwidth of 50 MHz, centred on 25 GHz. The reflectivity has been plotted assuming lossy dielectrics, with $\tan\delta\sim\,$few$\times 10^{-4}$. More specifically, we assume ${\rm Im}(n)=10^{-3}$. }
\label{fig:80disks1}
\end{center}
\end{figure}

As it took several hundred disks or higher harmonics for the transparent mode to exhibit significant distortion due to the non-zero velocity, we expect that the extrapolation to $80$ disks in~\cite{TheMADMAXWorkingGroup:2016hpc,Millar:2016cjp} should not be marred by the inclusion of 1D velocity effects.
To check this, we explicitly generate a broadband configuration using 80 dielectric disks, 1~mm thick with $n=5$, with a mirror on one side. Such a large number of disks has not been calculated in a realistic, optimised setup before, and allows us to check the extrapolations made in \cite{TheMADMAXWorkingGroup:2016hpc}. For consistency with~\cite{TheMADMAXWorkingGroup:2016hpc}, we will also use 25 GHz as our benchmark frequency, and optimise for a bandwidth of 50 MHz. The boost factor $\beta$, reflectivity $|{\cal R}|$ and group delay $\frac{\partial}{\partial \nu}\text {Arg}(\mathcal R)$ can be seen in figure~\ref{fig:80disks1}. This 80 disk setup gives a power boost ($\beta^2$) within 5\% of the prediction made in \cite{TheMADMAXWorkingGroup:2016hpc} using the Area Law (that is, that the area under $\beta^2(\nu)$ should increase linearly with the number of disks). Increasing the number of disks increases the number of boundary conditions that must be satisfied, leading to configurations containing a more ``rippling" structure than the 20 disk configuration of the same width shown in~\cite{TheMADMAXWorkingGroup:2016hpc}. 
Similar substructure can also be seen if one looks at the reflectivity of the setup. The group delay maps out resonances and reveals a similar rippling structure to $\beta$. In addition, assuming that the dielectrics are lossy, with ${\rm Im}(n)=10^{-3}$, $|{\cal R}|$ maps out resonant structure within the haloscope. To show this structure clearly, in figure~\ref{fig:80disks1} the losses were exaggerated an order of magnitude when compared to a realistic dielectric such as LaAlO$_3$ at low temperatures.

\begin{figure}[htbp]
\begin{center}
\includegraphics[width=9.5cm]{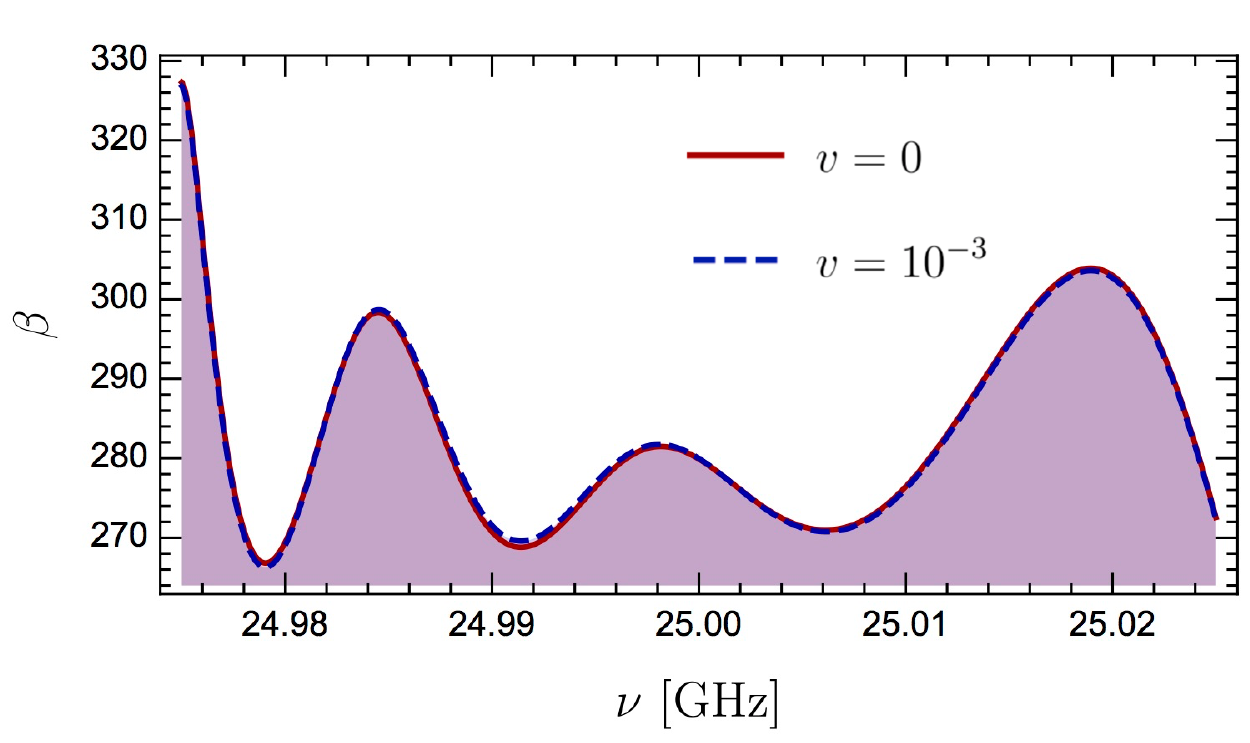}
\includegraphics[width=9.5cm]{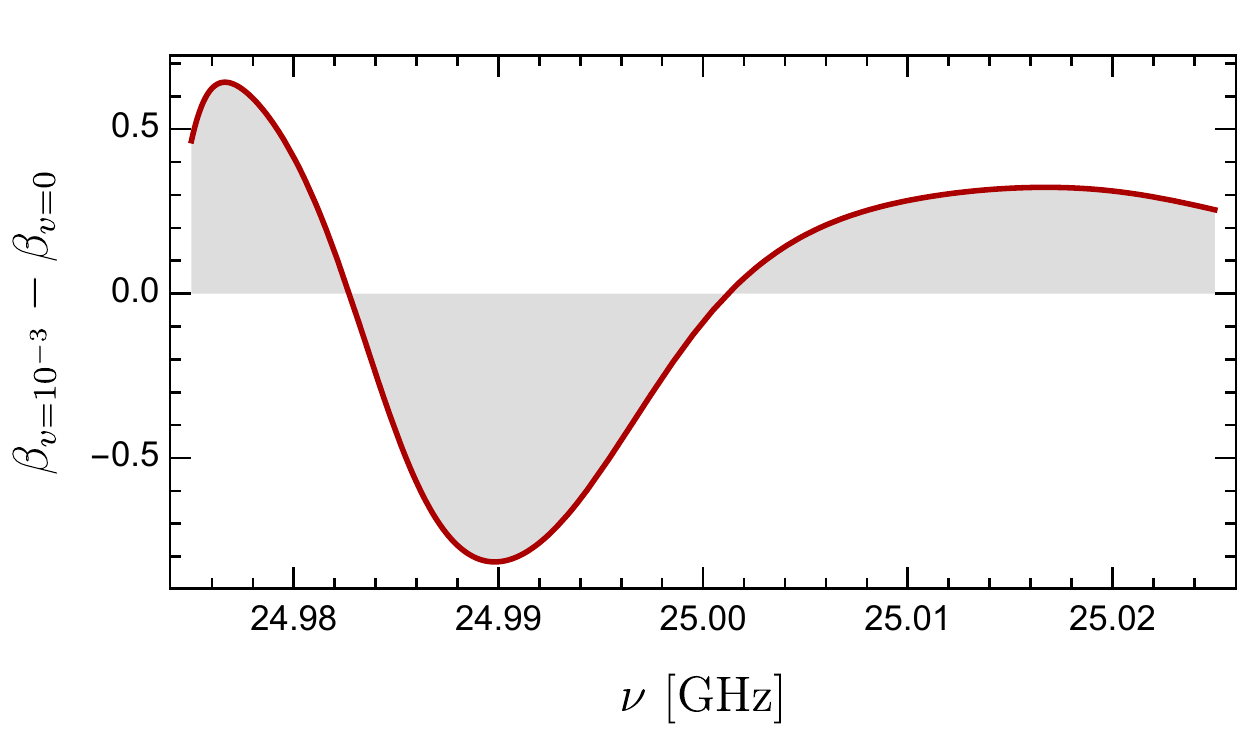}
\caption{{\it Top:}  Boost factor $\beta(\nu)$ for 80 disks optimised for a bandwidth of 50 MHz, centred on 25 GHz. Both $\beta_{v=0}$ and $\beta_{v=10^{-3}}$ are plotted as red and dotted blue curves, respectively. {\it Bottom:} Difference between $\beta$ for $v=10^{-3}$ and $v=0$ as a function of frequency. Inside the region of interest (the shown 50 MHz bandwidth), this difference corresponds to a less than percent level effect. }
\label{fig:80disks2}
\end{center}
\end{figure}

With such a disk configuration, we can directly see the effects of including a non-zero axion velocity. This is shown by plotting $\beta_{v=10^{-3}}-\beta_{v=0}$ as a function of frequency in the bottom panel of figure \ref{fig:80disks2}. The difference is entirely negligible, with less than a percent level change in the region of interest (as seen in the top panel of figure \ref{fig:80disks2}). 
This is actually somewhat less than one would expect from studying the transparent mode. As the spacing between each disk is close to $\lambda/2$, 80 disks in transparent mode (as shown in the top left panel of figure \ref{fig:Tdisks}) should be comparable with our 80 disk broadband configuration. 
However, the transparent setup is actually more prone to velocity effects than the realistic configurations. This is probably because of the large levels of symmetry present in the transparent mode; destructive and constructive effects occur coherently. That being said, the size of the distortion is similar enough to conclude that the scale at which velocity effects becomes important depends only weakly on the configuration as long as the total size of the haloscope remains constant. Thus dielectric haloscopes of the size considered in~\cite{TheMADMAXWorkingGroup:2016hpc,Millar:2016cjp} are free to neglect the 1D effects of a finite axion velocity without negative consequence. 

\begin{figure}[t]
\begin{center}
\includegraphics[width=10cm]{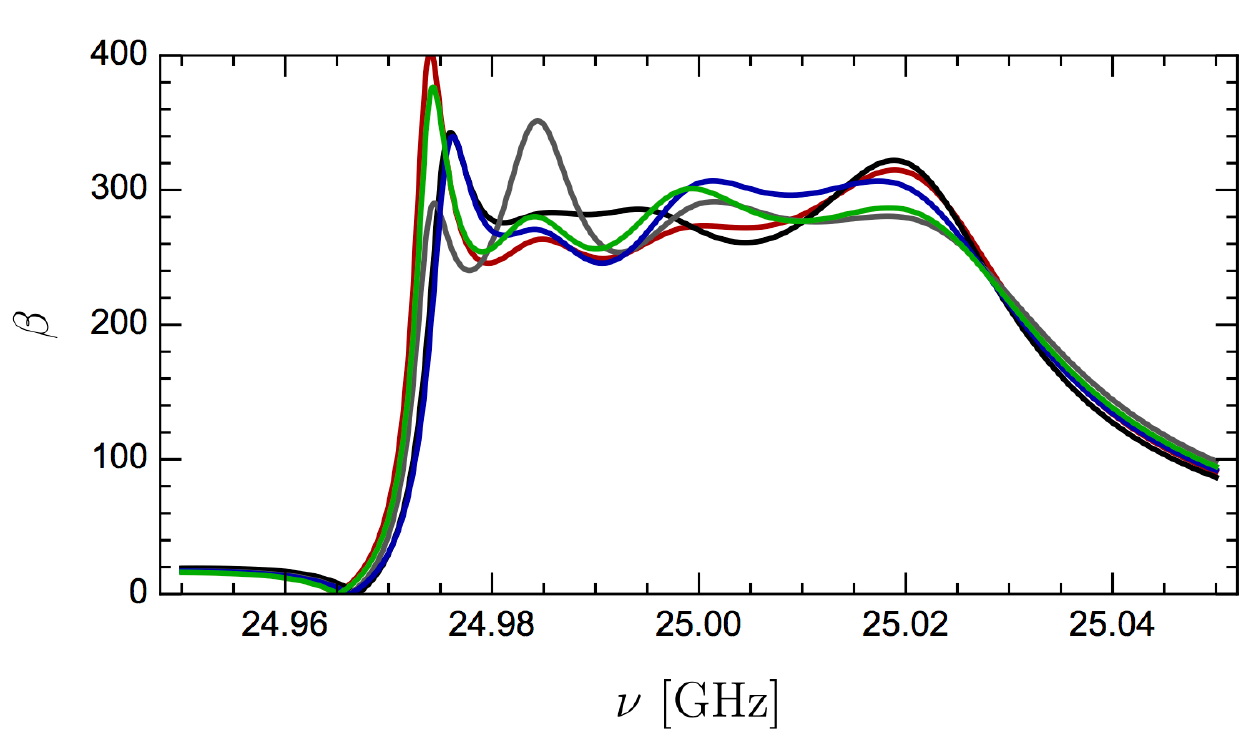}
\includegraphics[width=10cm]{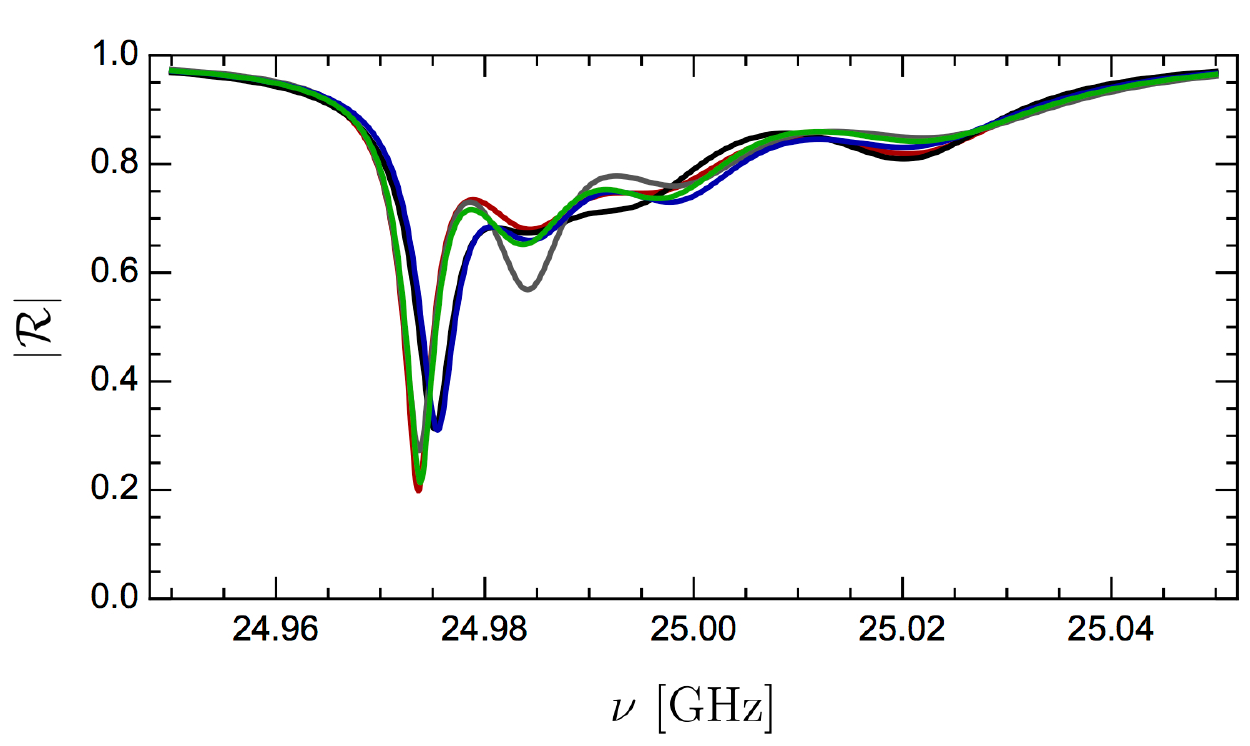}
\caption{Boost factor $\beta(\nu)$ (top) and reflectivity $|{\cal R}(\nu)|$ (bottom) for 80 disks optimised for a bandwidth of 25 MHz, centred on 25 GHz. Five configurations are shown, with each disk perturbed from the configuration shown in figure~\ref{fig:80disks1} by a top hat error function with width $5~\mu$m. For the reflectivity we assume losses exaggerated by an order of magnitude over the realistic value for visual comparison, with ${\rm Im}(n)=10^{-3}$. Losses are neglected for the boost factor.}
\label{fig:error}
\end{center}
\end{figure}

Note that errors in the disk placement made when readjusting the disks between measurements will likely result in larger changes to $\beta$ than the axion velocity. While the exact sensitivity to mispositioning errors depends on the distribution function of these errors, from numerical simulation we can see that one will require ${\cal O}(\mu{\rm m})$ placement position to keep the change in $\beta$ at the 10\% level. One can see an example assuming a top hat error function with a width of $5~\mu$m in figure~\ref{fig:error}. This gives a deviation in $\beta$ of order $10-20\%$, which should be an acceptable deviation for an experiment. Further, one can see that the reflectivity $|{\cal R}|$ remains correlated with $\beta$, allowing one to gain information about the deviation of $\beta$, allowing for errors to be corrected. For example, the large deviation of the gray curve can be seen in both the reflectivity and boost factor.  While the distortion from mispositioning errors is a greater effect than that of non-zero axion velocities, the time-dependence of changes to the axion velocity may allow it to be extracted from a signal.

\section{Probing the axion velocity with directional sensitivity}
\label{vdist}
Finally, we turn our attention to the benefits and drawbacks of velocity sensitivity. Whether or not the axion velocity plays a role depends on the type of experiment under consideration. Broadly speaking, there are two types of experimental setups: a search experiment, in which one is trying to verify the existence of DM axions, and an ``axion telescope", which would try to extract the most astrophysical information possible out a discovered DM signal.

\subsection{Search experiment} When searching for axions, the velocity effects must be small, i.e., $\beta(v)$ should be approximately independent of $v$. To see why we must require this, recall that the generated power at given axion momentum given by \eqref{eq:power} is 
\be
P({\bf p})\propto |a({\bf p})|^2\beta^2({\bf p}). \label{eq:power2}
\ee
Note $P({\bf p})$ is our only observable: as we do not know the velocity distribution of the axion $a({\bf p})$ a priori, one cannot disentangle $a({\bf p})$ from $\beta({\bf p})$, unless $\beta({\bf p})$ is independent of $v$. Consider if an experiment had a strongly ${\bf p}$ dependent $\beta$. When no signal is seen in a given measurement, we could not tell whether there were no axions present, or if there were DM axions with an unexpected velocity distribution giving a different boost factor than expected. In other words, the experiment does not distinguish between $a({\bf p})
\sim 0$ and $\beta({\bf p})\sim 0$, so no limits can be easily set. Thus, a design requirement for this type of setup is that $\beta(\vec p)= \beta(m_a)$.

As we saw in section~\ref{analytic}, velocity effects become relevant when the haloscope is on the order of $20\%$ of the axion de Broglie wavelength. This gives us a limit on the potential size of the device, corresponding to several hundred disks, as can be seen from considering the transparent mode in section \ref{transparent}. Ultimately, due to the Area Law found in~\cite{Millar:2016cjp}, this will provide a (very high) limit on the possible boost factor/bandwidth achievable by a search experiment type dielectric haloscope. 

Despite the lack of direct directional sensitivity, there is still a wealth of information that can be extracted after detection. When an axion signal is found, measuring $P(\omega)$ gives immediately the axion power spectrum $|a(\omega)|^2$ from \eqref{eq:power2}, as $\beta$ would be roughly constant and known by design. Note that $P$ only has a direct directional sensitivity if $\beta$ does. However, we could easily extract the mass and an angle averaged velocity distribution, 
\be
F(\omega) =  \langle |a(\vec p)|^2\rangle = 
\frac{v \omega^4}{2V\rho_{\rm a}} \int \frac{d\Omega}{(2\pi)^3} |a(\vec p)|^2. 
\ee 

With current receiver technology the number of channels in a search experiment can easily be $\gtrsim 10^4$; once a signal is detected these can be easily refocused to within the axion linewidth, allowing a very precise measurement of $F$. This alone contains significant astrophysical information: for example, if the axion DM forms a Bose-Einstein condensate as proposed in \cite{Sikivie:2009qn}, we would expect a Dirac delta power spectrum rather than say a Maxwell-Boltzmann spectrum, as is often assumed. One could also look for transient phenomena, though detecting substructure such as axion miniclusters~\cite{Hogan:1988mp} and associated tidal streams~\cite{Tinyakov:2015cgg} could take many years of measurement~\cite{OHare:2017yze}. If found these would give strong hints about the origin and structural formation of axion DM. Recall that we only measure the axion generated $E$-field (like any other haloscope experiment), which depends on both the coupling $g_{a\gamma}$ and the DM density $\rho_a$ given by \eqref{eq:density}. Because of this a haloscope experiment cannot determine the absolute local axion DM density and thus the normalisation of $a({\bf p})$. Breaking this degeneracy requires an independent measurement of $g_{a\gamma}$, an extraordinary challenge in this mass range.

Interestingly, $F(\omega)$ depends on time, due to the relative motion of the experiment with respect to the DM wind. With sufficiently long measurements one could detect annual and diurnal modulation, though the latter would be difficult~\cite{Turner:1990qx,OHare:2017yze}. The time dependence of $F(\omega)$ enters via very small shifts in the frequency of the axion, coming from small changes to the relative velocity of the DM with respect to the lab location on Earth. Annual modulation gives a relatively large ($5\%$) change to the velocity, but of course happens over yearly timescales~\cite{Turner:1990qx,Ling:2004aj}. By contrast diurnal modulation is relatively rapid, but so slight (on the order of 0.1\%~\cite{Turner:1990qx,Ling:2004aj}) that a very large amount of accumulated statistics would be needed to detect it.
These measurements will help to confirm the veracity of the signal and give information about the local structure of DM. This would give the rough direction of the DM wind, as well as potentially detecting some substructure in the halo.

\subsection{Axion telescope} Once we know the axion mass and $F(\omega)$, we could become much more ambitious and move towards an ``axion telescope". In this case, we desire the extra information contained in the axion velocity distribution, the full $a({\bf p})$. To extract this, we need an experiment in which the output power is sensitive to ${\bf p}$, i.e., we require $\beta$ to have a strong ${\bf p}$ dependence. 

To do this one would build an experiment with a size comparable to the axion de Broglie wavelength: as this is highly sensitive to $v_x$, we essentially measure the fraction of DM with a velocity in the $x$ direction (or a range of velocities, depending on the details of the setup). This is similar to the type of elongated cavity experiments proposed in \cite{Irastorza:2012jq}, which would also be sensitive to the change in phase of the axion over the device.
Using knowledge of the axion mass and power spectrum obtained from the search experiment, the dielectric disks could be placed to provide a very large signal, allowing for shorter measurements. For example, we could measure for a few hours rather than a day or a week as is currently considered~\cite{MADMAXinterestGroup:2017bgn}. These shorter measurements would be sensitive to velocity changes on shorter time scales, giving a more rapid accumulation of statistics.\footnote{Note that one does not actually have to preform shorter measurements, rather we simply want that the statistics accumulate at a faster rate. One could do a single long measurement and break the data down into different times during analysis.}

While the rotation of the Earth only changes the magnitude of the axion velocity by 0.1\%, the changing direction gives a sizeable modulation of $v_x$ depending on the alignment of the $x$-direction of the experiment and the DM wind. This directional modulation of the axion velocity can be used to probe the axion velocity structure in a 2D plane.\footnote{This 2D plane is given by the circle traced out by the $x$ direction of the experiment as the Earth rotates. Note that this circle changes over the course of the year, allowing for limited 3D coverage.} As the device would now be the same size as the de Broglie wavelength, there would be an ${\cal O}(1)$ variation in the signal power, sensitive to much smaller variations in $v_x$ than the ${\cal O}(10^{-9})$ shift in the axion frequency would allow. This would give a much more detailed map of the local DM than is possible with directionally insensitive devices, hence the name ``axion telescope".
We could build further evidence for the astrophysical origin of the signal by comparing very precisely the observed directionality of the signal with theory (i.e., the expected DM wind). Unlike a search experiment, any features observed would come with directional information; a search experiment can only infer the direction of DM wind using knowledge of the motion of the Earth. For example, if we see features such as axion miniclusters or tidal streams~\cite{OHare:2017yze} we would know the direction that these objects came from. 
When combined with simulations of galaxy formation this directional information could be a test of structure formation. To aid in this, it might be useful to also have a velocity-insensitive device operating next to the axion telescope, giving a simultaneous measurement of the overall axion power spectrum and the $x$-moving component of DM at that time. Alternatively, one could use multiple experiments pointed in three directions to have a complete 3D local map at any given point in time.

\begin{figure}[t]
\begin{center}
\includegraphics[width=9.5cm]{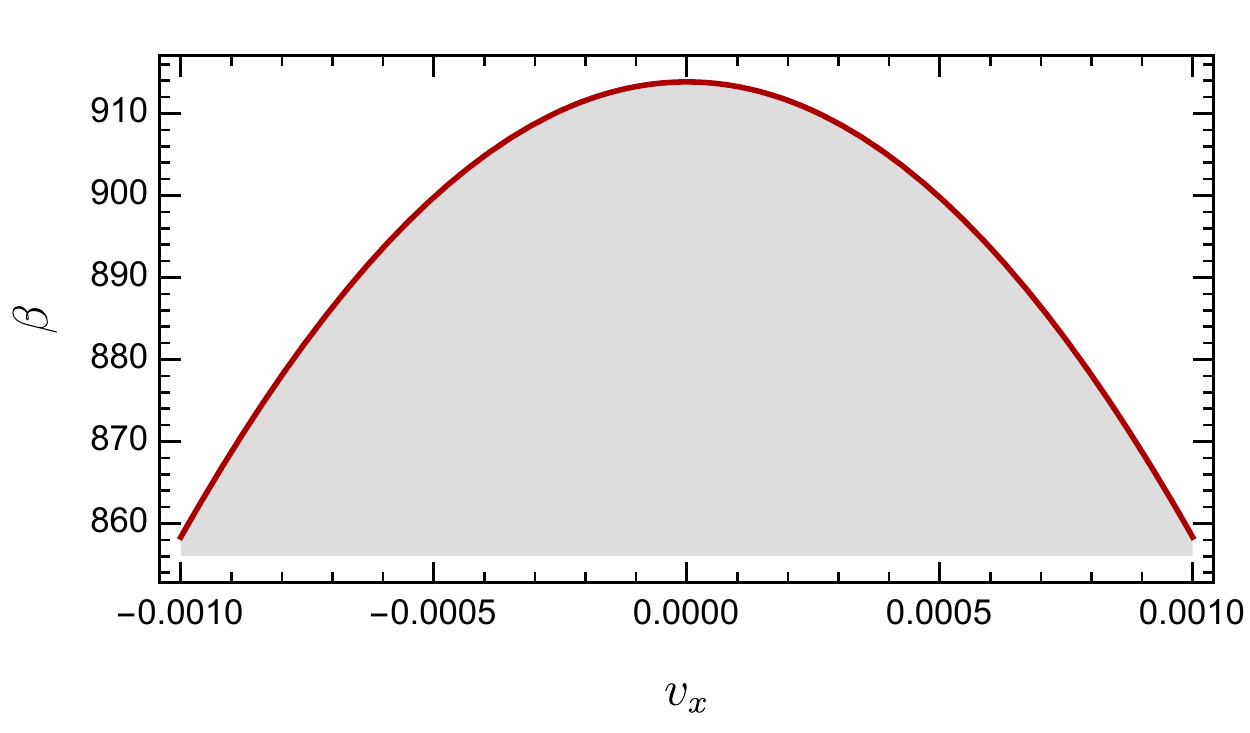}
\caption{Boost factor $\beta(v_x)$ for 80 disks of 1~mm thickness optimised for a 25 GHz axion (with a bandwidth of 1 MHz). The disk spacings have been optimised around $5\lambda /2$. This gives the rough scale of diurnal modulations due to the changing axion velocity. Here we have assumed that $a({\bf p})$ is a delta function. }
\label{fig:80disks3}
\end{center}
\end{figure}

Directional sensitivity can be partially achieved even with the search experiment. If the axion were discovered at the high mass end of the scanned range, as the corresponding Compton wavelength $\lambda$ is small there would be a lot of unused magnet volume. As the spacings between the dielectrics would be roughly $\lambda/2$ during a regular search~\cite{Millar:2016cjp}, as $\lambda$ decreases the total distance taken up by the haloscope decreases. We could then increase the spacing between the disks to a higher harmonic configuration where the distance is $\sim 3\lambda/2, 5\lambda/2$, etc. instead of $\sim \lambda/2$. In figure \ref{fig:80disks3} we show the $v_x$ dependence for a configuration optimised for a 25 GHz axion signal, using spacings of order $5\lambda/2$. For illustrative purposes, 
we assumed a Dirac delta velocity distribution, and have neglected the effects caused by the ${\cal O}(10^{-9})$ change in frequency, i.e., we show only the boost factor dependence. Note that the exact distribution does not significantly change this result: if a Maxwell-Boltzmann distribution is used instead the effect of the directional change is similar. While the exact velocity variation during the Earth's rotation depends on the location and alignment of the experiment, figure~\ref{fig:80disks3} gives the rough scale of the boost factor modulation. Maximal diurnal modulations can be achieved by aligning the $x$-direction of the experiment with the DM wind at some time of the day: this point gives the maximum $v_x$, with the minimum occurring 12 hours later. 
The power modulation caused by the axion velocity is typically 10\%, significantly more pronounced than for the configuration shown in figures~\ref{fig:80disks1} and \ref{fig:80disks2}, and potentially detectable.

For most cases the effects of the DM velocity are slight for an experiment of a similar size to the planned MADMAX~\cite{MADMAXinterestGroup:2017bgn}, but in the event that axions are discovered these effects contain a wealth of information about the DM in our universe, and potentially the galactic structure. As dielectric haloscopes are sensitive not only to the overall velocity distribution, but in principal have directional sensitivity, they would be a very powerful tool to study the DM in our galaxy.

\section{Conclusions and outlook}
\label{conclusion}
The axion is one of the best motivated DM candidates, potentially resolving both the DM question as well as the strong CP problem. The parameter space in which axions can form the DM of our Universe can be heavily restricted by theoretical considerations; in particular, high mass axions in the range $50-200~\mu$eV are predicted in scenarios where the PQ symmetry is broken after inflation. Dielectric haloscopes seem to be a promising avenue for searching for these axions, allowing one to use a detection device with a much larger volume than a traditional resonant cavity. 

The axion's non-zero velocity can in principal have non-trivial effects in devices where the linear dimensions are comparable to the axion de Broglie wavelength. To study this in detail, we derived the axion-photon mixing equations with a non-zero velocity, and then looked at the photon production due to changing media. From this, we saw that the main influence of the velocity of the axion comes from the change of phase of the axion itself over the haloscope: effects from the transverse velocities and axion induced $H$-field were always subdominant. With a generalised transfer matrix formalism, we could then study the change to the boost factor in detail. We found that for haloscopes smaller than $\sim15 - 20$\% of the axion de Broglie wavelength, the axions velocity could be safely ignored. 

To see this in realistic setting, we studied a realistic 80 disk setup for the first time. While in normal situations velocity effects were entirely negligible, if one increased the spacing between each disk by odd multiples of $\lambda/2$ it is possible to see a noticeable shift in the boost factor. This could be used in the event of a potential discovery of the axion, allowing one to observe effects such as diurnal modulation. This would be both a useful check on the veracity of the signal, as well as allow one to preform astrophysical measurements, such as testing for diurnal modulation with directional sensitivity.

\section*{Acknowledgments}

We thank the MADMAX Working Group at the Max Planck Institute for Physics
for support and helpful discussions, in particular Georg Raffelt. We would also like to thank Edoardo Vitagliano for helpful discussions and Jan Schuette-Engel for pointing out typos and a sign error in section 3. We acknowledge partial support by the
Deutsche Forschungsgemeinschaft through Grant No.\ EXC 153 (Excellence
Cluster ``Universe'') and Grant No.\ SFB 1258
(Collaborative Research Center ``Neutrinos, Dark Matter, Messengers''), as well as the European Union through the Initial
Training Network ``Elusives'' Grant No.\ H2020-MSCA-ITN-2015/674896 and Grant No.\
H2020-MSCA-RISE-2015/690575 (Research and Innovation Staff Exchange project ``Invisibles Plus''). J.R.\ is supported by the Ramon y Cajal Fellowship 2012-10597 and FPA2015-65745-P (MINECO/FEDER).  Part of this work was performed at the Bethe Forum ``Axions'' (7--18 March 2016), Bethe Center for Theoretical Physics, University of Bonn, Germany.

\clearpage



\begin{thebibliography}{99}

\bibitem{Peccei:2006as}
  R.~D.~Peccei,
  The Strong CP problem and axions,
  Lect.\ Notes Phys.\  {\bf 741} (2008) 3
  [arXiv:0607268].

\bibitem{Kim:2008hd}
  J.~E.~Kim and G.~Carosi,
  Axions and the strong CP problem,
  Rev.\ Mod.\ Phys.\  {\bf 82} (2010) 557
  [arXiv:0807.3125].

\bibitem{Agashe:2014kda}
  K.~A.~Olive {\it et al.} [Particle Data Group Collaboration],
  Review of Particle Physics,
  Chin.\ Phys.\ C {\bf 38} (2014) 090001.

\bibitem{Sikivie:2009fv}
  P.~Sikivie,
  Dark matter axions,
  Int.\ J.\ Mod.\ Phys.\ A {\bf 25} (2010) 554
  [arXiv:0909.0949].

\bibitem{Kawasaki:2013ae}
  M.~Kawasaki and K.~Nakayama,
  Axions: Theory and cosmological role,
  Ann.\ Rev.\ Nucl.\ Part.\ Sci.\  {\bf 63} (2013) 69
  [arXiv:1301.1123].

\bibitem{Hiramatsu:2012gg}
  T.~Hiramatsu, M.~Kawasaki, K.~Saikawa and T.~Sekiguchi,
  Production of dark matter axions from collapse of string-wall systems,
  Phys.\ Rev.\ D {\bf 85} (2012) 105020;
  Erratum {\em ibid.} {\bf 86} (2012) 089902
  [arXiv:1202.5851].

\bibitem{Kawasaki:2014sqa}
  M.~Kawasaki, K.~Saikawa and T.~Sekiguchi,
  Axion dark matter from topological defects,
  Phys.\ Rev.\ D {\bf 91} (2015) 065014
  [arXiv:1412.0789].
 

  
  
\bibitem{TheMADMAXWorkingGroup:2016hpc}
  A.~Caldwell {\it et al.} [MADMAX Working Group],
 Dielectric haloscopes: a new way to detect axion dark matter,
  Phys.\ Rev.\ Lett.\  {\bf 118} (2017)  091801
  [arXiv:1611.05865].
\bibitem{Sikivie:1983ip}
  P.~Sikivie,
  Experimental tests of the invisible axion,
  Phys.\ Rev.\ Lett.\  {\bf 51} (1983) 1415;
  Erratum {\em ibid.} {\bf 52} (1984) 695.

\bibitem{Rybka:2014xca}
  G.~Rybka [ADMX Collaboration],
  Direct detection searches for axion dark matter,
  Phys.\ Dark Univ.\  {\bf 4} (2014) 14.
\bibitem{Kenany:2016tta}
  S.~Al Kenany {\it et al.},
 Design and operational experience of a microwave cavity axion detector for the 20–100 $\mu$eV range,
  Nucl.\ Instrum.\ Meth.\ A {\bf 854} (2017) 11
  [arXiv:1611.07123].
  
\bibitem{Rybka:2014cya}
  G.~Rybka, A.~Wagner, A.~Brill, K.~Ramos, R.~Percival and K.~Patel,
Search for dark matter axions with the Orpheus experiment,
  Phys.\ Rev.\ D {\bf 91} (2015)  011701
  [arXiv:1403.3121].
\bibitem{Goryachev:2017wpw}
  M.~Goryachev, B.~T.~Mcallister and M.~E.~Tobar,
  Axion detection with cavity arrays,
  arXiv:1703.07207.
\bibitem{McAllister:2017ern}
  B.~T.~McAllister, G.~Flower, L.~E.~Tobar and M.~E.~Tobar,
  Tunable super-mode dielectric resonators for axion haloscopes,
  arXiv:1705.06028.
\bibitem{McAllister:2017lkb}
  B.~T.~McAllister, G.~Flower, J.~Kruger, E.~N.~Ivanov, M.~Goryachev, J.~Bourhill and M.~E.~Tobar,
 The ORGAN experiment: an axion haloscope above 15 GHz,
  arXiv:1706.00209.
  
\bibitem{Horns:2012jf}
  D.~Horns, J.~Jaeckel, A.~Lindner, A.~Lobanov, J.~Redondo and A.~Ringwald,
  Searching for WISPy cold dark matter with a dish antenna,
  JCAP {\bf 1304} (2013) 016
  [arXiv:1212.2970].

  
  \bibitem{Millar:2016cjp}
  A.~J.~Millar, G.~G.~Raffelt, J.~Redondo and F.~D.~Steffen,
  Dielectric haloscopes to search for axion dark matter: theoretical foundations,
  JCAP {\bf 1701} (2017) 061
  [arXiv:1612.07057].
  
\bibitem{Ioannisian:2017srr}
  A.~N.~Ioannisian, N.~Kazarian, A.~J.~Millar and G.~G.~Raffelt,
 Axion-photon conversion caused by dielectric interfaces: quantum field calculation,
  arXiv:1707.00701.



  
\bibitem{diCortona:2015ldu}
  G.~Grilli di Cortona, E.~Hardy, J.~P.~Vega and G.~Villadoro,
  The QCD axion precisely,
  JHEP {\bf 1601} (2016) 034
  [arXiv:1511.02867].

\bibitem{Kim:1979if}
  J.~E.~Kim,
  Weak interaction singlet and strong CP invariance,
  Phys.\ Rev.\ Lett.\  {\bf 43} (1979) 103.

\bibitem{Shifman:1979if}
  M.~A.~Shifman, A.~I.~Vainshtein and V.~I.~Zakharov,
  Can confinement ensure natural CP invariance of strong interactions?,
  Nucl.\ Phys.\ B {\bf 166} (1980) 493.

\bibitem{Dine:1981rt}
  M.~Dine, W.~Fischler and M.~Srednicki,
  A simple solution to the strong CP problem with a harmless axion,
  Phys.\ Lett.\ B {\bf 104} (1981) 199.

\bibitem{Zhitnitsky:1980tq}
  A.~R.~Zhitnitsky,
  On possible suppression of the axion hadron interactions,
  Yad.\ Fiz.\  {\bf 31} (1980) 497
  [Sov.\ J.\ Nucl.\ Phys.\  {\bf 31} (1980) 260].

\bibitem{Kim:2014rza}
  J.~E.~Kim,
  Calculation of axion–-photon–-photon coupling in string theory,
  Phys.\ Lett.\ B {\bf 735} (2014) 95; Erratum {\em ibid.} {\bf 741} (2014) 327
  [arXiv:1405.6175].
\bibitem{DiLuzio:2016sbl}
  L.~Di Luzio, F.~Mescia and E.~Nardi,
  Redefining the axion window,
  Phys.\ Rev.\ Lett.\  {\bf 118} (2017),  031801
  [arXiv:1610.07593].
\bibitem{DiLuzio:2017pfr}
  L.~Di Luzio, F.~Mescia and E.~Nardi,
 The window for preferred axion models,
  arXiv:1705.05370.
\bibitem{Wilczek:1987mv}
  F.~Wilczek,
  Two applications of axion electrodynamics
  Phys.\ Rev.\ Lett.\  {\bf 58} (1987) 1799.
  \bibitem{Das:2004qka}
  S.~Das, P.~Jain, J.~P.~Ralston and R.~Saha,
  JCAP {\bf 0506} (2005) 002
  [arXiv:0408198].

\bibitem{Das:2004ee}
  S.~Das, P.~Jain, J.~P.~Ralston and R.~Saha,
 The dynamical mixing of light and pseudoscalar fields
  Pramana {\bf 70} (2008) 439
  [arXiv:0410006].
\bibitem{Ganguly:2008kh}
  A.~K.~Ganguly, P.~Jain and S.~Mandal,
  Photon and axion oscillation in a magnetized medium: A general treatment,
  Phys.\ Rev.\ D {\bf 79} (2009) 115014
  [arXiv:0810.4380].
  
  
\bibitem{VISINELLI:2013fia}
  L.~Visinelli,
 Axion-electromagnetic waves,
  Mod.\ Phys.\ Lett.\ A {\bf 28} (2013),  1350162
  [arXiv:1401.0709].
\bibitem{Raffelt:1987im}
  G.~Raffelt and L.~Stodolsky,
  Mixing of the photon with low mass particles,
  Phys.\ Rev.\ D {\bf 37} (1988) 1237.

  
  
\bibitem{Dobrynina:2014qba}
  A.~Dobrynina, A.~Kartavtsev and G.~Raffelt,
  Photon-photon dispersion of TeV gamma rays and its role for
  photon-ALP conversion,
  Phys.\ Rev.\ D {\bf 91} (2015) 083003
  [arXiv:1412.4777].
  
  \bibitem{Galan:2015msa}
  J.~Galán {\it et al.},
 Exploring 0.1–10 eV axions with a new helioscope concept,
  JCAP {\bf 1512} (2015) 012
  [arXiv:1508.03006].
  
\bibitem{Sikivie:2013laa}
  P.~Sikivie, N.~Sullivan and D.~B.~Tanner,
  Proposal for axion dark matter detection using an LC circuit,
  Phys.\ Rev.\ Lett.\  {\bf 112} (2014)  131301
  [arXiv:1310.8545 ].


\bibitem{Catena:2009mf}
  R.~Catena and P.~Ullio,
  A novel determination of the local dark matter density,
  JCAP {\bf 1008} (2010) 004
  [arXiv:0907.0018].
\bibitem{Strigari:2009zb}
  L.~E.~Strigari and R.~Trotta,
  Reconstructing WIMP properties in direct detection experiments including
  Galactic dark matter distribution uncertainties,
  JCAP {\bf 0911} (2009) 019
  [arXiv:0906.5361].

\bibitem{Weber:2009pt}
  M.~Weber and W.~de Boer,
  Determination of the local dark matter density in our Galaxy,
  Astron.\ Astrophys.\  {\bf 509} (2010) A25
  [arXiv:0910.4272].

\bibitem{Bovy:2012tw}
  J.~Bovy and S.~Tremaine,
  On the local dark matter density,
  Astrophys.\ J.\  {\bf 756} (2012) 89
  [arXiv:1205.4033].

\bibitem{Nesti:2013uwa}
  F.~Nesti and P.~Salucci,
  The Dark Matter halo of the  Milky Way, AD 2013,
  JCAP {\bf 1307} (2013) 016
  [arXiv:1304.5127].

\bibitem{Bozorgnia:2013pua}
  N.~Bozorgnia, R.~Catena and T.~Schwetz,
  Anisotropic dark matter distribution functions and impact on WIMP direct detection,
  JCAP {\bf 1312} (2013) 050
  [arXiv:1310.0468].

\bibitem{Read:2014qva}
  J.~I.~Read,
  The Local Dark Matter Density,
  J.\ Phys.\ G {\bf 41} (2014) 063101
  [arXiv:1404.1938].

  
\bibitem{Redondo:2010js}
  J.~Redondo,
  Photon-axion conversions in transversely inhomogeneous magnetic fields,
  arXiv:1003.0410.
\bibitem{Jaeckel:2013sqa}
  J.~Jaeckel and J.~Redondo,
  An  antenna for directional detection of WISPy dark matter,
  JCAP {\bf 1311} (2013) 016
  [arXiv:1307.7181].
\bibitem{Jaeckel:2015kea}
  J.~Jaeckel and S.~Knirck,
  Directional resolution of dish antenna experiments to search for WISPy dark matter,
  JCAP {\bf 1601} (2016) 005
  [arXiv:1509.00371].
  
  \bibitem{Irastorza:2012jq}
  I.~G.~Irastorza and J.~A.~Garcia,
  Direct detection of dark matter axions with directional sensitivity,
  JCAP {\bf 1210} (2012) 022
  [arXiv:1207.6129].



\bibitem{Krauss:1985ub}
  L.~Krauss, J.~Moody, F.~Wilczek and D.~E.~Morris,
  Calculations for cosmic axion detection,
  Phys.\ Rev.\ Lett.\  {\bf 55} (1985) 1797.
  
  
\bibitem{Hong:2014vua}
  J.~Hong, J.~E.~Kim, S.~Nam and Y.~Semertzidis,
  Calculations of resonance enhancement factor in axion-search tube-experiments,
  arXiv:1403.1576 .
  
\bibitem{Sikivie:2009qn}
  P.~Sikivie and Q.~Yang,
  Bose-Einstein condensation of dark matter axions,
  Phys.\ Rev.\ Lett.\  {\bf 103} (2009) 111301
  [arXiv:0901.1106].
  
  \bibitem{Hogan:1988mp}
  C.~J.~Hogan and M.~J.~Rees,
  Phys.\ Lett.\ B {\bf 205} (1988) 228.
\bibitem{Tinyakov:2015cgg}
  P.~Tinyakov, I.~Tkachev and K.~Zioutas,
  JCAP {\bf 1601} (2016)  035
  [arXiv:1512.02884].
  
   \bibitem{OHare:2017yze}
  C.~A.~J.~O'Hare and A.~M.~Green,
 Axion astronomy with microwave cavity experiments,
  Phys.\ Rev.\ D {\bf 95} (2017)  063017
  [arXiv:1701.03118].
\bibitem{Turner:1990qx}
  M.~S.~Turner,
  Periodic signatures for the detection of cosmic axions,
  Phys.\ Rev.\ D {\bf 42} (1990) 3572.
 
  
\bibitem{Ling:2004aj}
  F.~S.~Ling, P.~Sikivie and S.~Wick,
 Diurnal and annual modulation of cold dark matter signals,
  Phys.\ Rev.\ D {\bf 70} (2004) 123503
  [astro-ph/0405231].



%
\bibitem{MADMAXinterestGroup:2017bgn}
  P.~Brun {\it et al.} [MADMAX Interest Group],
  A new experimental approach to probe QCD Axion Dark Matter in the mass range above 40 µeV, https://www.mpp.mpg.de/fileadmin/user\_upload/Forschung/MADMAX/madmax\_white\_paper.pdf.

\end{thebibliography}
\end{document}